\documentclass[prd,superscriptaddress,amsfonts,amssymb,amsmath,showpacs,twocolumn]{revtex4}
\usepackage{bm}
\usepackage{amsfonts}
\usepackage{latexsym}
\usepackage[latin1]{inputenc}
\usepackage{graphicx}
\usepackage{amsmath}
\usepackage{palatino}
\usepackage{mathpazo}
\usepackage{textcomp}
\usepackage{float}
\usepackage{booktabs}
\usepackage{dcolumn}
\usepackage{hyperref}
\hypersetup{colorlinks,citecolor=blue}
\usepackage{amsmath}
\usepackage{xcolor}
\usepackage{mathtools}
\usepackage{orcidlink}
\usepackage[caption=false]{subfig}
\usepackage{commath}
\def\jnl@style{\it}
\def\aaref@jnl#1{{\jnl@style#1}}

\def\aaref@jnl#1{{\jnl@style#1}}

\def\aj{\aaref@jnl{AJ}}                   
\def\apj{\aaref@jnl{ApJ}}                 
\def\apjl{\aaref@jnl{ApJ}}                
\def\apjs{\aaref@jnl{ApJS}}               
\def\apss{\aaref@jnl{Ap\&SS}}             
\def\aap{\aaref@jnl{A\&A}}                
\def\aapr{\aaref@jnl{A\&A~Rev.}}          
\def\aaps{\aaref@jnl{A\&AS}}              
\def\mnras{\aaref@jnl{Mon.~Not.~Roy.~Astron.~Soc.}}             
\def\prd{\aaref@jnl{Phys.~Rev.~D}}        
\def\prc{\aaref@jnl{Phys.~Rev.~C}}  
\def\prl{\aaref@jnl{Phys.~Rev.~Lett.}}    
\def\qjras{\aaref@jnl{QJRAS}}             
\def\skytel{\aaref@jnl{S\&T}}             
\def\ssr{\aaref@jnl{Space~Sci.~Rev.}}     
\def\zap{\aaref@jnl{ZAp}}                 
\def\nat{\aaref@jnl{Nature}}              
\def\aplett{\aaref@jnl{Astrophys.~Lett.}} 
\def\apspr{\aaref@jnl{Astrophys.~Space~Phys.~Res.}} 
\def\physrep{\aaref@jnl{Phys.~Rep.}}      
\def\physscr{\aaref@jnl{Phys.~Scr}}       
\def\commat{\aaref@jnl{Comm.~Math.~Phys.}}              
\def\science{\aaref@jnl{Science}}               
\def\cqg{\aaref@jnl{Classical Quant.~Grav.}}            
\def\jpcs{\aaref@jnl{JPCS}}                                     
\def\ijmpd{\aaref@jnl{Int.~J.~Mod.~Phys.~D}}                    
\def\grg{\aaref@jnl{Gen.~Relat.~Gravit.}}               
\def\rpp{\aaref@jnl{Rep.~Prog.~Phys.}}          
\def\npa{\aaref@jnl{Nucl.~Phys.~A}}        
\def\lrr{\aaref@jnl{Living Rev.~Rel.}}                   
\def\jcap{\aaref@jnl{J.~Cosmology Astropart.~Phys.}}    
\def\rmp{\aaref@jnl{Rev.~Mod.~Phys.}}   

\allowdisplaybreaks[1]

\begin{document}
\color{red}

\title{Effective equation of state in modified gravity and observational
constraints}

\author{Simran Arora\orcidlink{0000-0003-0326-8945}}
\email{dawrasimran27@gmail.com}
\affiliation{Department of Mathematics, Birla Institute of Technology and
Science-Pilani, \\Hyderabad Campus, Hyderabad-500078, India.}
\author{Xin-he Meng\orcidlink{0000-0001-7681-6063}}
\email{xhm@nankai.edu.cn}
\affiliation{School of Physics, Nankai Univ, P.R. China.}
\author{S. K. J. Pacif\orcidlink{0000-0003-0951-414X}}
\email{shibesh.math@gmail.com}
\affiliation{Department of Mathematics, School of Advanced Sciences, Vellore
Institute of Technology, Vellore 632014, Tamil Nadu, India.}
\author{P.K. Sahoo\orcidlink{0000-0003-2130-8832}}
\email{pksahoo@hyderabad.bits-pilani.ac.in}
\affiliation{Department of Mathematics, Birla Institute of Technology and
Science-Pilani, \\Hyderabad Campus, Hyderabad-500078, India.}
\date{\today}

\begin{abstract}
In this article, the bulk viscosity is introduced in a modified gravity
model. The gravitational action has a general $f(R,T)$ form, where $R$ and $%
T $ are the curvature scalar and the trace of energy momentum tensor
respectively. An effective equation of state (EoS) has been investigated in
the cosmological evolution with bulk viscosity. In the present scenario, the
Hubble parameter which has a scaling relation with the redshift can be
obtained generically. The role of deceleration parameter $q$ and equation of
state parameter $\omega $ is discussed to explain the late-time accelerating
expansion of the universe. The statefinder parameters and Om diagnostic
analysis are discussed for our obtained model to distinguish from other dark
energy models together with the analysis of energy conditions and velocity
of sound for the model. We have also numerically investigated the model by
detailed maximum likelihood analysis of $580$ Type Ia supernovae from Union $%
2.1$ compilation datasets and updated $57$ Hubble datasets ($31$ data points
from differential age method and $26$ points from BAO and other methods). It
is with efforts found that the present model is in good agreement with
observations.
\end{abstract}

\maketitle



\section{Introduction}\label{I}

We still believe in that the Einstein's general theory of relativity (GR)
does not give the final word to all gravity phenomena, though in the solar
system tests GR is very successful so far. As well understood that after
many observational and experimental tests, there are some issues which hint
towards a possible modification to general theory of relativity (GR) at
large scales such as the cosmic accelerating expansion phenomena and dark
matter mysteries. The recently developed late time accelerated expansion 
\cite{1,2,3} of the Universe is an immediate motivation for the same. The
simplest possible modification for such an acceleration is to consider a
cosmological constant $\Lambda $ existence which plays the role of dark
energy, i.e. the fluid responsible for an effective negative pressure.
Another way to identify the role of dark energy is to treat it as an
effective geometrical quantity coming out of a modified Einstein-Hilbert
action. We can do this by replacing the Ricci curvature $R$ in the Einstein
action by a generic function $f(R)$ which gives rise to the named $f(R)$
theories as mentioned in \cite{4,5,Yousaf3,Yousaf4}, for example.

Several studies have been carried out on the modified theories of gravity
which can explain both early and late time expansion of the universe. The $%
f(G)$ gravity form, for example, is also an important modified theory of
gravity in which there is replacement of $R$ by a general function $f(G)$,
where $G$ is the Gauss-Bonnet invariant \cite{6,7}. Other modified theories
of gravity including $f(T)$ form where $T$ is the torsion, $f(R,G)$, and
also $f(R,T)$ forms, etc. without by direct introducing an effective dark
energy term (cosmological constant like), can also give a satisfying
explanation to present cosmic acceleration expansion. Some related works on
these theories have been described in refs. \cite%
{8,9,11,Yousaf1,Yousaf2,Pedro}.

\textbf{Viscosity}: In order to portray the recent accelerated expansion
era, the framework of GR and so called cosmological constant $\Lambda $CDM
model with vacuum and dust energy is not sufficient as it is faced with some
shortcomings. The two main issues are the coincidence puzzle and the fine
tuning problem. Though the evolutions of dark matter and dark energy are
different, they are faced with the coincidence densities. On the other hand
the fine tuning problem is associated with the disparity between the
theoretical and the observational value of the cosmological constant. These
problems have provoked the deliberations of various dark energy models like
quintessence, perfect fluid models, scalar fields. Apart from these many
authors have stated that the cosmic viscosity directs the late time
acceleration expansion. The viscosity theories in cosmology is important
when connected with the early universe, i.e. when the temperature was about $%
10^{4}$K (at the time of neutrino parting). There are two different
viscosity coefficients in cosmic fluid namely bulk viscosity $\zeta $ and
shear viscosity $\eta $. We omit shear viscosity due to the accepted spatial
isotropy of the universe like the Robertson-Walker metric descriptions.

By considering a bulk viscous fluid the problem of finding a viable
mechanism for the origin of bulk viscosity in the expanding universe arises.
Theoretically, bulk viscosity exists due to the deviations from the local
thermodynamic irreversibility of the motion. In cosmology, bulk viscosity
arises as an effective pressure to restore the system back into its thermal
equilibrium\cite{Okumura}. Eckart \cite{Eckart} made the first approach for
describing non-equilibrium thermodynamic effects in a relativistic context.
It has also been studied that the bulk viscosity is sufficient to drive the
cosmic fluid from the quintessence to phantom region \cite{Brevik}. Sharif
and Yousaf \cite{Yousaf} have also investigated into stability regions for a
non-static restricted class of axially symmetric geometry. The work includes
shearing viscous fluid that collapse non-adiabatically.

\textbf{Viscosity in modified gravity}: There have been a great variety of
models describing the universe with dark energy discussed above. But if we
talk about the problem of cosmic adaptation, i.e. the mean stage of low
redshift, the cosmic accelerating expansion can be justified by the approach
of modification in Einstein equations geometrically. Bulk viscosity can also
produce an acceleration without the need of scalar field or cosmological
constant if connected to inflation. The bulk viscosity contributes to the
pressure term and exerts extra pressure driving the accelerating expansion
of the universe \cite{20}. Also, the effective negative pressure due to the
viscous media effects the key condition to generate inflation.

Most of the time argument on standard gravity assume the cosmic fluid to be
ideal that is non-viscous. If we see from hydrodynamics point of view, two
viscosity coefficients discussed above come into play which means deviation
from thermal equilibrium to the first order. This theory is an acceptance of
Eckart 1940 theory. The important part of this is the non-casual behavior.
Therefore taking second order deviations from a thermal equilibrium leads to
a casual theory respecting special relativity. Now, it is also important to
take into account some more realistic models, which process due to
complicated viscosity, as that Singh and Kumar \cite{10} have studied the
role of bulk viscosity in the evolution of the Universe by considering the
modified $f(R,T)$ gravity model. There were remarkable cosmological
applications of viscous imperfect fluids in 1970s \cite{33,34}. Also many
other authors have investigated the idea of bulk viscous fluids to explain
the acceleration of the Universe expansion \cite{12,27,35,36,Brevik}. Davood 
\cite{Davood} also studied the role of bulk viscosity in $f(T)$ gravity. The
cosmic pressure in this phenomenon is considered as $p=(\gamma -1)\rho
-3\zeta H$, where the $\gamma $ parameterizes the EoS \cite{29}. The form of
this pressure was originally proposed by Eckart \cite{Eckart}. However,
Eckart theory undergoes some anatomies. One of those is the instability of
the equilibrium states \cite{Hiscock}. Another is that the dissipative
perturbations propagate at infinite speeds \cite{W. Israel}. In 1979, a more
general theory was developed by Israel and Stewart \cite{Stewart} which was
casual and stable. Eckart theory can also be obtained in the first order
limit in Stewart theory when the relaxation time tends to zero. So, if we
talk about the limiting case, Eckart theory is a good approximation which is
also discussed in \cite{Titus}. Hence, we know that Eckart theory is less
complicated than the Israel-Stewart theory irrespective of drawbacks it
have. Many authors have also pointed out that, the relaxation time is to be
constant in Israel-Stewart theory which is not reasonably correct in
expanding universe.

From the observational constraints, we have that the current EoS parameter $%
\omega =\frac{p}{\rho }$ is around $-1$ \cite{22,30}, probably larger than $%
-1$ by the recent DES results, which is called the quintessence range while
the EoS below $-1$ corresponding to the so called phantom region. In this
article we have observed that the focused model we have investigated into
shows accelerating behavior and behaves as the quintessence alike ($\omega
>-1$) as current datasets favored.

Fisher and Carlson \cite{Fisher} have examined the form of $f(R,T)= f_{1}(R)
+ f_{2}(T)$, in which they state that $f(R,T)$ yields a new physics and
limits could be placed on the cross-terms by comparison with observations.
This work is again reexamined by Harko and Moreas \cite{Tiberiu Harko} in
examining observational restrictions on the function $f_{2}(T)$. Setare and
Houndjo \cite{Setare} have studied the finite-time future singularities
model in $f(T)$ gravity with the effect of viscosity. Sharif and Rani \cite%
{Rani} have also worked on viscous dark energy in $f(T)$ gravity. The work
of Iver Brevik \cite{Iver} describes viscosity in $f(R)$ gravity.

In the present article, we study the Friedman-Lem\^aitre-Robertson-Walker
(FLRW) geometric frame model with bulk viscosity effects in the modified $%
f(R,T)$ gravity theory, in which we have investigated into a general
effective equation of state form given by, \newline
$p=(\gamma -1)\rho +p_{0}+\omega_{H}H+\omega _{H2}H^{2}+\omega _{dH}\dot{H}$%
, \newline
and also shown that the following time-dependent bulk viscosity \newline
$\zeta =\zeta _{0}+\zeta _{1}H+\zeta_{2}(\frac{\dot{H}}{H}+H)$ \newline
is the same form as derived by the above mentioned effective EoS. The field
equations and its exact solutions are obtained with constant $\alpha $ by
assuming the model simplest form of $f(R,T)=R+2f(T)$ where $f(T)=\alpha T$.
This is the simplest functional choice of $f(R,T)$ gravity as when $\alpha =
0$, the field equations correspond to that of GR ones.

The article has been discussed in various sections as follows. In section %
\ref{II}, we have formulated the field equations followed by modeling with
viscosity. We have described the general solution and the behavior of
various parameters in section \ref{III}. In Sec \ref{IV} we have performed
various tests to check the validation of model containing the energy
conditions, velocity of sound, statefinder parameter, the Om diagnostics and
also observational datasets corresponding to SNeIa and $H(z)$. And the last
section \ref{V} is followed by the conclusion. We have taken the Einstein
field equations in units of $8\pi G=c=1$.

\section{Field equations}\label{II}

The $f(R,T)$ theory is a modified theory of gravity in which the most
general action for $f(R,T)$ gravity is given as in Ref. \cite{11, 10}

\begin{equation}  \label{e1}
S=\frac{1}{2}\int d^{4}x\sqrt{-g}(f(R,T)+ 2L_{m}),
\end{equation}
where the Einstein-Hilbert Lagrangian, $R$, has been replaced by an
arbitrary function of the Ricci scalar curvature $R$ and the trace $T$ of
the energy momentum tensor. Here, $g$ is the determinant of the metric
tensor $g_{\mu \nu}$ and $L_{m}$ is the matter Lagrangian density.

Variation of the action with respect to the metric tensor gives us the
following gravitational field equation:

\begin{multline}  \label{e2}
f_{R}(R,T)R_{\mu \nu}-\frac{1}{2} f(R,T)g_{\mu \nu}+(g_{\mu
\nu}\square-\nabla_{\mu} \nabla_{\nu}) f_{R}(R,T)= \\
T_{\mu \nu}-f_{T}(R,T)T_{\mu \nu}-f_{T}(R,T)\Theta_{\mu \nu},
\end{multline}
where, $\nabla _{\mu }$ and $\nabla _{\nu }$ represents the covariant
derivative and $\Theta _{\mu \nu }$ is defined by 
\begin{equation}
\Theta _{\mu \nu }\equiv g^{\alpha \beta }\frac{\delta T_{\alpha \beta }}{%
\delta g^{\mu \nu }}.  \label{e3}
\end{equation}

We consider the Friedman-Lemaitre-Robertson-Walker (FLRW) metric in the flat
space geometry ($k=0$) 
\begin{equation}
ds^{2}=dt^{2}-a^{2}(t)[dr^{2}+r^{2}d\theta ^{2}+r^{2}\sin ^{2}\theta d\phi
^{2}]  \label{e4}
\end{equation}%
where $a(t)$ is the cosmic scale factor.

The components of four-velocity $u^{\mu}$ are $u^{\mu}=(1,0)$ in comoving
coordinates. Assume that the cosmic fluid possesses a bulk viscosity $\zeta$%
. We have the energy-momentum tensor for a viscous fluid as follows 
\begin{equation}  \label{e5}
T_{\mu \nu}= \rho u_{\mu}u_{\nu}-\overline{p} h_{\mu \nu}.
\end{equation}
where $h_{\mu \nu}= g_{\mu \nu}+ u_{\mu}u_{\nu}$ and $\overline{p}= p-3\zeta
H$ is the effective pressure.

If we choose the Lagrangian density as $L_{m}= -\overline{p}$ then the
tensor $\Theta_{\mu \nu}$ becomes 
\begin{equation}  \label{e6}
\Theta_{\mu \nu}= -2T_{\mu \nu}-\overline{p} g_{\mu \nu}.
\end{equation}
Using \eqref{e5} and \eqref{e6}, the field equation for the bulk viscous
fluid become 
\begin{equation}  \label{e7}
R_{\mu \nu}-\frac{1}{2}Rg_{\mu \nu}= T_{\mu \nu}+ 2f^{\prime
}(T)T_{\mu\nu}+(2 \overline{p}f^{\prime }(T)+f(T)) g_{\mu\nu}.
\end{equation}

For the particular choice of the function $f(T)=\alpha T$, where $\alpha $
is a constant, we get field equations as 
\begin{equation}
3H^{2}=\rho +2\alpha (\rho +\overline{p})+\alpha T,  \label{e8}
\end{equation}%
\begin{equation}
2\dot{H}+3H^{2}=-\overline{p}+\alpha T,  \label{e9}
\end{equation}%
where $T=\rho -3\overline{p}$. From Eqs. \eqref{e8} and \eqref{e9}, we have 
\begin{equation}
\ 2\dot{H}+(1+2\alpha )(p+\rho )-3(1+2\alpha )\zeta H=0.  \label{e10}
\end{equation}

\section{General Solution}\label{III}

We can see the Eqs. \eqref{e8} and \eqref{e9} contains four unknown
parameters viz. $\rho, p, \zeta \& H$. To get an exact solution, we need two
more physically viable equations. As discussed in the introduction, we shall
consider the following EoS (an explicit form as given in \cite{29}) 
\begin{equation}
p=(\gamma -1)\rho +p_{0}+\omega _{H}H+\omega _{H2}H^{2}+\omega _{dH}\dot{H},
\label{e11}
\end{equation}
where $p_{0},\omega _{H},\omega _{H2},\omega _{dH}$ are free parameters. If
we compare with the bulk viscosity form we get the most general one. We show
that this time-dependent bulk viscosity 
\begin{equation}
\zeta =\zeta _{0}+\zeta _{1}\frac{\dot{a}}{a}+\zeta _{2}\frac{\ddot{a}}{\dot{%
a}},  \label{e12}
\end{equation}
is effectively equivalent to the form derived by using Eq. \eqref{e11} where 
$\zeta _{0},\zeta _{1},\zeta _{2}$ are constants.

The reason behind this is 
\begin{align*}
\overline{p}= p-3\zeta H &= p-3 (\zeta_{0}+\zeta_{1} \frac{\dot{a}}{a}
+\zeta_{2} \frac{\ddot{a}}{\dot{a}}) H \\
&= p-3 \zeta_{0} H-3 \zeta_{1} H^{2}-3 \zeta_{2}(\dot{H}+H^{2})
\end{align*}
which gives 
\begin{equation}  \label{e13}
\overline{p}= p-3 \zeta_{0} H-3(\zeta_{1}+\zeta_{2}) H^{2}-3 \zeta_{2} \dot{H%
}.
\end{equation}

We can obtain the corresponding coefficients are 
\begin{align*}
& \omega_{H} =-3 \zeta_{0}, \\
& \omega_{H2} =-3 (\zeta_{1} + \zeta_{2}), \\
& \omega_{dH}=-3 \zeta_{2}.
\end{align*}

Using Eqs. \eqref{e8}, \eqref{e11}, \eqref{e12}, \eqref{e13}, we obtain the
explicit form of energy density as 
\begin{equation}
\rho =\frac{\alpha p_{0}+2\alpha \omega _{H}H+(2\alpha
\omega_{H2}+3)H^{2}+2\alpha \omega _{dH}\dot{H}}{1+4\alpha -\alpha \gamma }.
\label{e14}
\end{equation}

Subsequently, using Eq.\eqref{e9} we obtain the bulk viscous pressure as

\begin{widetext}
\begin{equation}
\overline{p}=\frac{\alpha ^{2}p_{0}+2\alpha ^{2}\omega _{H}H+(2\alpha
^{2}\omega _{H2}-9\alpha -3+3\alpha \gamma )H^{2}+(2\alpha ^{2}\omega_{dH}-2(1+4\alpha -\alpha \gamma ))\dot{H}}{(1+4\alpha -\alpha \gamma)(1+3\alpha )}.  \label{e15}
\end{equation}

Using Eqs. \eqref{e10}, \eqref{e11}, \eqref{e14}, we have an equation 
\begin{multline}
\left[ 2+\frac{2(1+2\alpha )\omega _{dH}(1+4\alpha )}{1+4\alpha -\alpha\gamma }\right] \dot{H}+\left[ \frac{2(1+2\alpha )\omega _{H}(1+4\alpha )}{1+4\alpha -\alpha \gamma }\right] H+\left[ \frac{(1+2\alpha )(2\omega_{H2}(1+4\alpha )+3\gamma )}{1+4\alpha -\alpha \gamma }\right] H^{2}+
\\
\left[ \frac{(1+2\alpha )p_{0}(1+4\alpha )}{1+4\alpha -\alpha \gamma }\right]
=0.\label{e16} 
\end{multline}%
\end{widetext}

Because of the high non linearity, it is difficult to solve the above
equation \eqref{e16} for which without the loss of generality we assume that 
$p_{0}=0$. This simplifies the equation to give the time evolution of Hubble
parameter $H$ as, 
\begin{equation}
H=\frac{k_{1}}{k_{3}e^{k_{1}t}-k_{2}},  \label{e17}
\end{equation}%
where $k_{1}=\dfrac{\frac{2(1+2\alpha )\omega _{H}(1+4\alpha )}{%
1+4\alpha-\alpha \gamma }}{2+\frac{2(1+2\alpha )\omega _{dH}(1+4\alpha )}{%
1+4\alpha-\alpha \gamma }}$, $k_{2}=\dfrac{\frac{(1+2\alpha )(2\omega
_{H2}(1+4\alpha)+3\gamma )}{1+4\alpha -\alpha \gamma }}{2+\frac{2(1+2\alpha
)\omega_{dH}(1+4\alpha )}{1+4\alpha -\alpha \gamma }}$ and $k_{3}=k_{1}c_{1}$%
, with $c_{1}$ being a constant of integration.

Using the definition $H=\frac{\dot{a}}{a}$, we can obtain the scale factor
given by, 
\begin{equation}  \label{e18}
a=k_{4} k_{3}^{-\frac{1}{k_{2}}}(k_{3}-k_{2} e^{-k_{1}t})^{\frac{1}{k_{2}}},
\end{equation}
where $k_{4}$ is a constant of integration.

Finally, the deceleration parameter ($q=-\frac{a\ddot{a}}{\dot{a}^{2}}$) is
obtained as, 
\begin{equation}
q=-1+k_{3}e^{k_{1}t}.  \label{e19}
\end{equation}%
Now, we have the general set of solution for the formulated system. In order
to discuss the detailed evolution of the Universe in various phases, we
shall discuss the behavior of the different cosmological parameters obtained
here. In the present article, we are interested to examine the different
regimes of the Universe particularly the phase transition from decelerated
to accelerated by constraining the model parameters. We know a positive
value of $q$ refers the decelerating phase while a negative value of $q$
corresponds to accelerating phase of the universe. So, we shall write all
the cosmological parameters in terms of redshift $z$ using the relation 
\begin{equation}
a(t)=\frac{1}{1+z}  \label{e20}
\end{equation}%
with $a_{0}=1$. The Hubble parameter and the deceleration parameter are two
observable parameters which can be rewritten in terms of redshift as, 
\begin{equation}
H(z)=\frac{k_{1}}{k_{2}}((k_{4}+k_{4}z)^{k_{2}}-1),  \label{e21}
\end{equation}%
and 
\begin{equation}
q(z)=-1+k_{2}\dfrac{(k_{4}+k_{4}z)^{k_{2}}}{(k_{4}+k_{4}z)^{k_{2}}-1}.
\label{e22}
\end{equation}%
Recent studies reveals that the present observed deceleration rate of the
Universe is $q_{0}=-0.51_{-0.01}^{+0.09}$ \cite{26} and a transition
redshift from deceleration to acceleration is $z_{t}=0.65_{-0.17}^{+0.19}$ 
\cite{32}. In literature \cite{21,23,24} reported that the Universe passed
from a decelerated phase to an accelerated one at $z_{t}\approx 0.7$ \cite%
{25} . So, we have chosen the values of these free parameters ($k_{1}$, $%
k_{2}$, $k_{3}$, $k_{4}$) in the present model in such way that our $q_{0}$
and $z_{t}$ result are consistent with values reported in the literature.
Henceforth, we will discuss a particular model as an exemplification and
study the cosmic history of the universe with some numerical choice of the
values of the model parameters. However, we have chosen the values of $k_{2}$
and $k_{4}$ that has been constrained from some observational datasets in
the subsequent section. The evolution of $q(z)$ is shown in the following
Fig. \ref{fig1} with suitable choice of the model parameters.

\begin{figure}[H]
\centering
\includegraphics[scale =0.4]{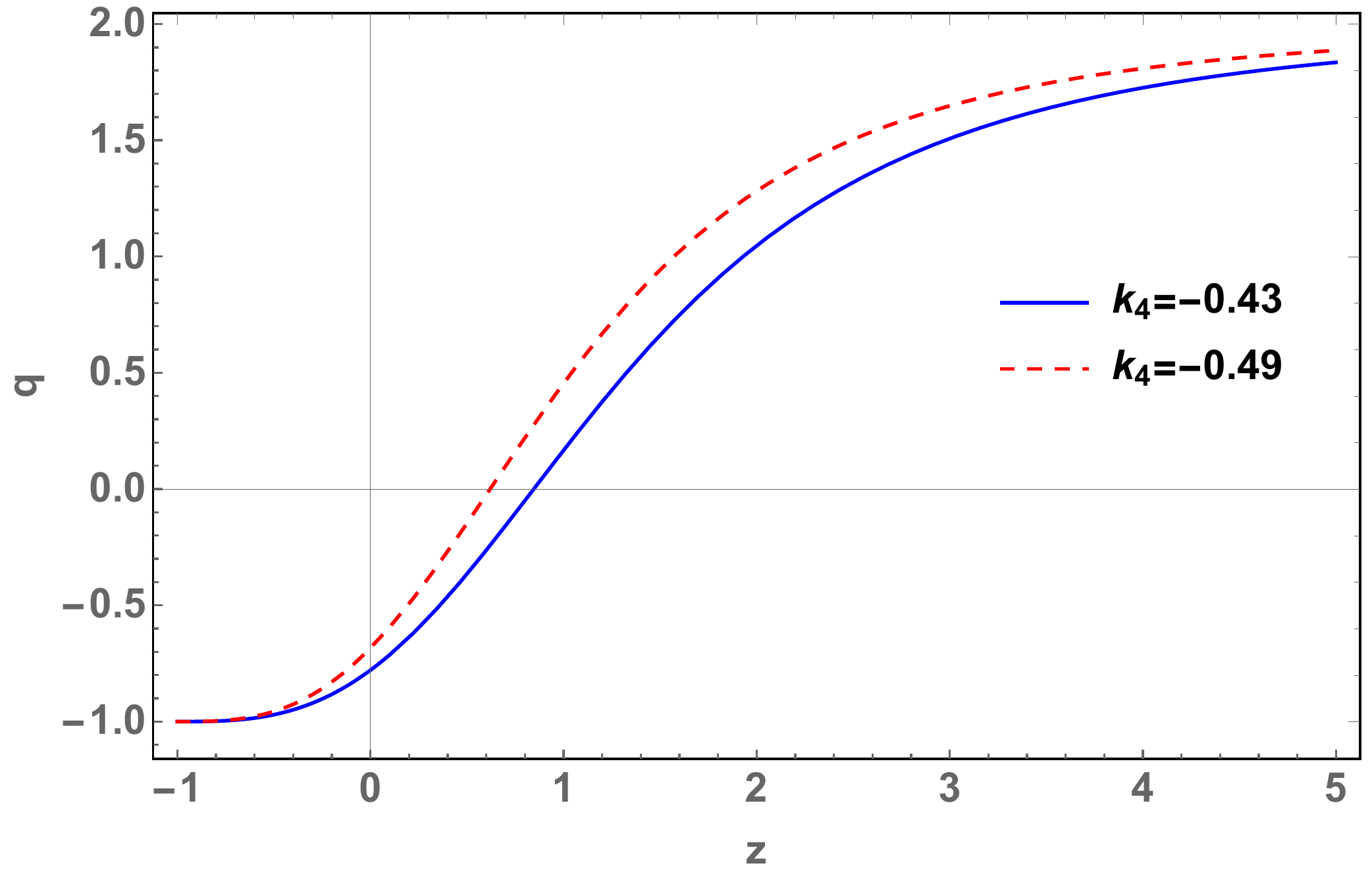}
\caption{The plot discusses the behavior of deceleration parameter versus
redshift $z$ for the model with $k_{2}=3$ and $k_{4}=-0.43, -0.49$.}
\label{fig1}
\end{figure}

From Fig. \ref{fig1}, we see that the deceleration parameter $q$ varies from
negative to positive at $z_{t}=0.845815$ and $z_{t}=0.619797$ with $%
q_{0}=-0.779046$ and $q_{0}=-0.684206$ with two different values of $%
k_{4}=-0.43$ and $k_{4}=-0.49$ respectively. This indicates, the universe
exhibits a transition from early deceleration to the current acceleration in
this model. The behavior of $\rho $ and $\bar{p}$ from equations \eqref{e14}
and \eqref{e15} with respect to redshift $z$ is plotted below.

\begin{figure}[ ]
\centering
\includegraphics[scale =0.3]{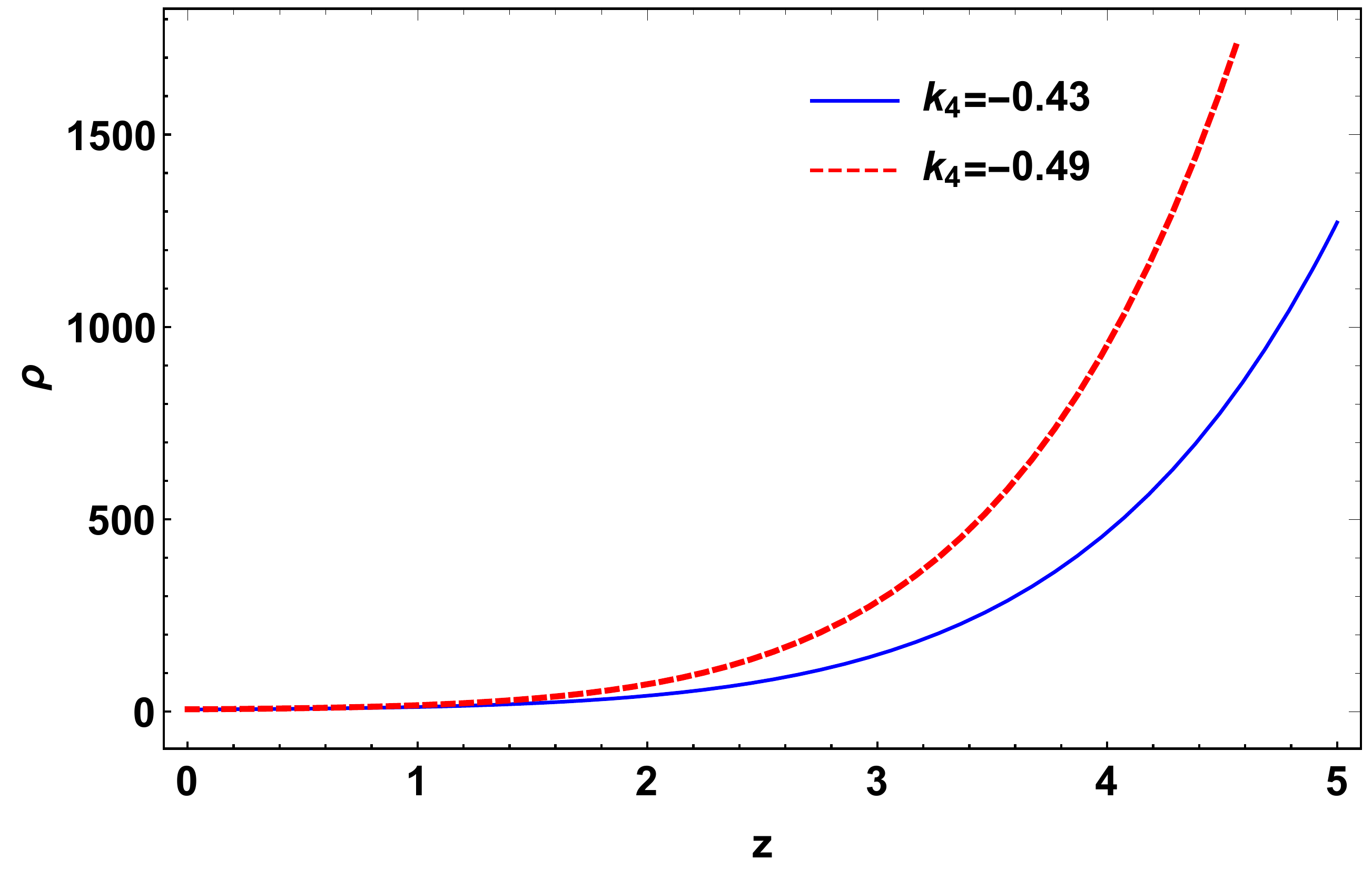}
\caption{The plot shows the behavior of density parameter of the model
versus redshift $z$ with $\protect\alpha=-0.1$, $\protect\gamma=1.01$, $%
\protect\omega_{H}=4.1$, $\protect\omega_{H2}=1.57$, $\protect\omega%
_{dH}=-0.1$ and $k_{4}=-0.43, -0.49$.}
\label{fig2}
\end{figure}

\begin{figure}[ ]
\centering
\includegraphics[scale =0.3]{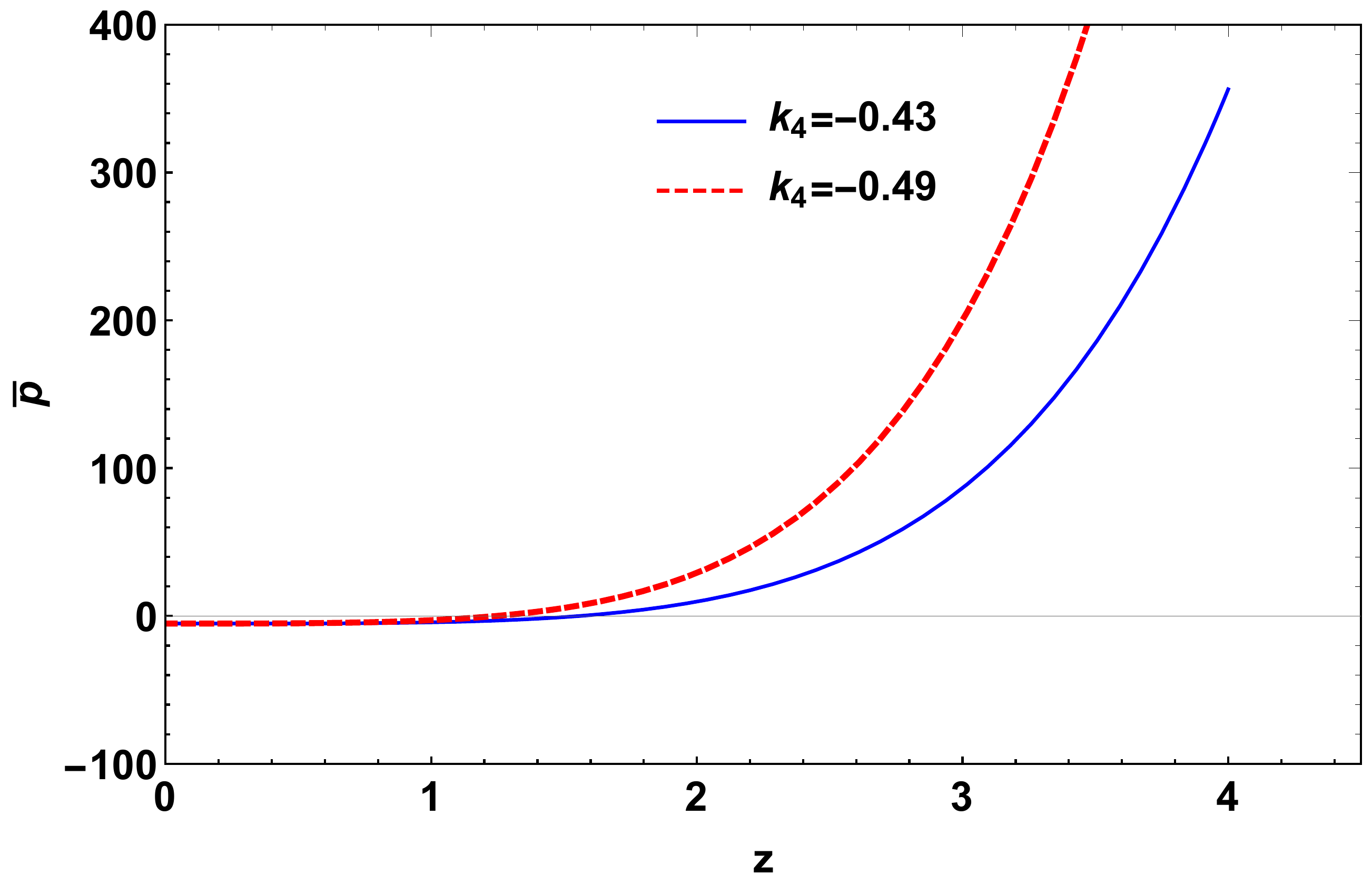}
\caption{The plot shows the behavior of effective pressure of the model
versus redshift $z$ with $\protect\alpha=-0.1$, $\protect\gamma=1.01$, $%
\protect\omega_{H}=4.1$, $\protect\omega_{H2}=1.57$, $\protect\omega%
_{dH}=-0.1$ and $k_{4}=-0.43, -0.49$.}
\label{fig3}
\end{figure}

Since we have constrained the values of $k_{2}$ and $k_{4}$ in section \ref%
{F}, so accordingly the values of other model parameters such as $\alpha $, $%
\gamma $, $\omega _{H}$, $\omega _{H2}$, $\omega _{dH}$ involved in $k_{1}$, 
$k_{2}$ and $k_{3}$ are set for the analysis. We can clearly observe the
behavior of $\rho $ and $\overline{p}$ from the Figs. \ref{fig2} \& \ref%
{fig3}, which shows that energy density is an increasing function of $z$ and
the effective pressure has a transition from negative to positive. The
present study demonstrate the expanding behavior of the universe and on the
other hand negative pressure indicates the cosmic accelerated expansion of
the universe.

\begin{figure}[tbp]
\centering
\includegraphics[scale =0.3]{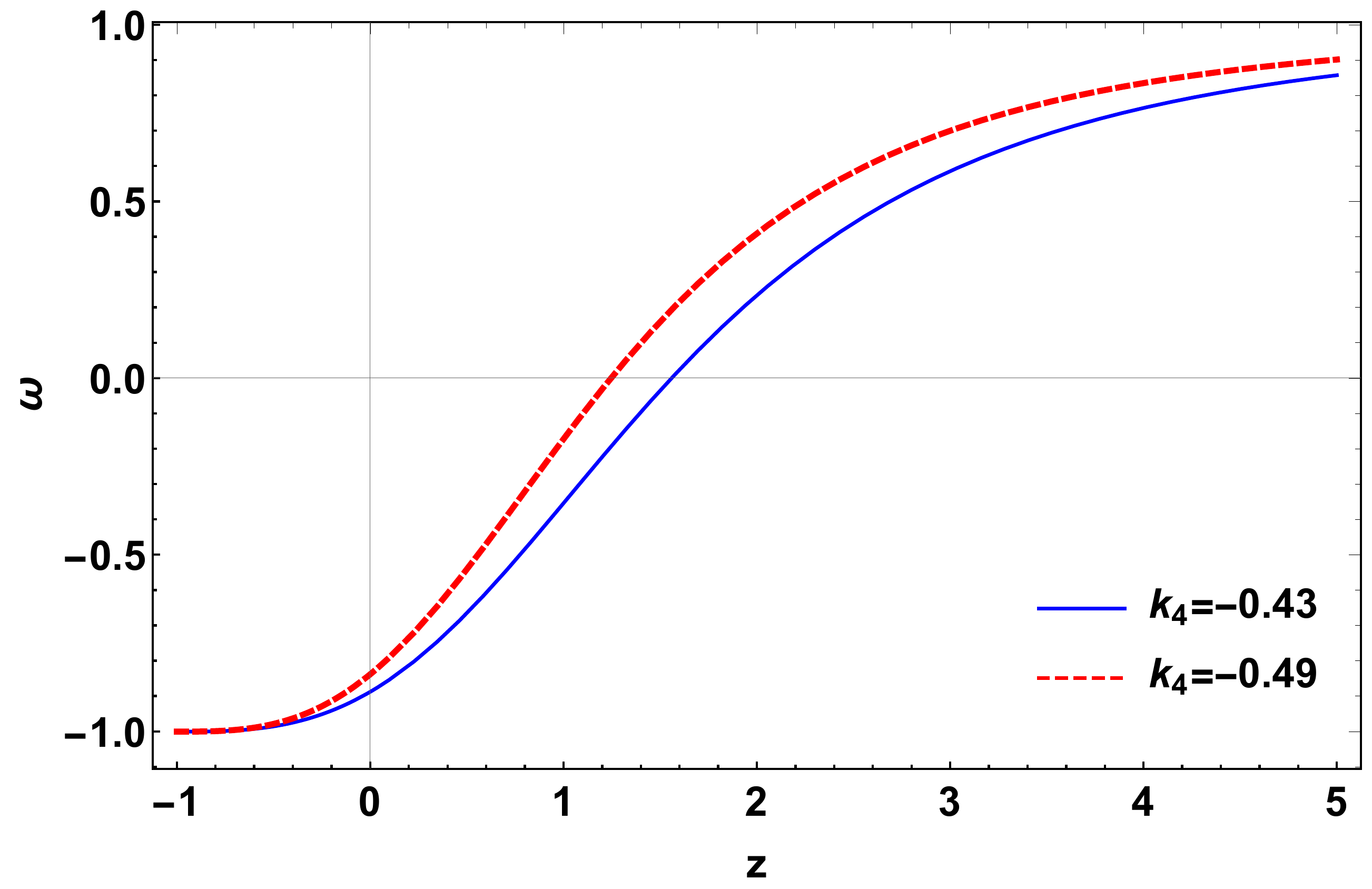}
\caption{The plot shows the behavior of Equation of State of the model
versus redshift $z$ with $\protect\alpha =-0.1$, $\protect\gamma =1.01$, $%
\protect\omega _{H}=4.1$, $\protect\omega _{H2}=1.57$, $\protect\omega %
_{dH}=-0.1$, $k_{4}=-0.43, -0.49$.}
\label{fig4}
\end{figure}

The EoS parameter is the relationship between pressure $p$ and energy
density $\rho $. The EoS parameter is used to classify the decelerated and
accelerated expansion of the universe and it categorizes various epochs as
follows: when $\omega =1$, it represents stiff fluid, if $\omega =1/3$, the
model shows the radiation dominated phase while $\omega =0$ represents
matter dominated phase. In the present accelerated phase of evolution, $%
0\geq \omega >-1$ shows the quintessence phase and $\omega =-1$ shows the
cosmological constant, i.e., $\Lambda $CDM model and $\omega <-1$ yields the
phantom era. In Fig. \ref{fig4}, we have plotted the EoS parameter versus
redshift $z$ by considering same values of the model parameters as discussed
above,

The graph in Fig. \ref{fig4} shows that the as $z\rightarrow -1$ , $\omega
\rightarrow -1$ in the future. It also shows the transition from negative to
positive in due course of evolution which indicates the earlier decelerating
phase of the universe with positive pressure (suitable for structure
formation) and present accelerating phase of the evolution with negative
pressure. The present values of the EoS parameter can be calculated as of $%
\omega _{0}=-0.888046$ for $k_{4}=-0.43$ and $\omega _{0}=-0.838394$ for $%
k_{4}=-0.49$ together with other stated values of other model parameters. In
the following section we discuss the of the obtained model with some
mathematical tools and observational datasets.

\section{Tests for validation of the model}\label{IV}

There are some theoretical and observational tests to check the validity of
any cosmological model. So now, we shall discuss some of the cosmological
tests for the validation of our obtained model.

\subsection{Energy conditions}

The energy conditions (ECs) of GR permit one to deduce very powerful and
general theorems about the behavior of strong gravitational fields and
cosmological geometries \cite{visser/2000}. ECs have a great adequacy in
classical GR which consider the singularity problems of space-time and
explain the behavior of null, space-like, time-like or light-like geodesics.
It provides us some flexibility to analyze certain ideas about the nature of
cosmological geometries and some relations that the stress energy momentum
must satisfy to make energy density positive. It is normally used in GR to
show and study the singularities of space-time \cite{wald}. In general, ECs
can be classified as a) SEC (Strong energy condition), b) DEC (Dominant
energy condition), c) WEC (Weak energy condition) and d) NEC (Null energy
condition) \cite{Hawking/1973}. The formulation of these four types of ECs
in GR is expressed as:

a) SEC: Gravity should always be attractive and in cosmology $\rho + 3p \geq
0$.

b) DEC: The matter energy density measured by any observer must be positive
and propagate in a causal way, which leads to $\rho \geq \mid p \mid $.

c) WEC: The matter energy density measured by any observer should be
positive, $\rho$ $\geq 0 $, $\rho+p \geq 0 $.

d) NEC: It's the minimum requirement that is implied by SEC and WEC, is $%
\rho+p \geq 0$.

The violation of NEC implies that none of the mentioned ECs are validated.
The SEC is currently the subject of much discussion for the current
accelerated expansion of the Universe \cite{Barcelo/2002, moraes/2017}. SEC
must be violated in cosmological scenarios during the inflationary expansion
and at the present time \cite{Visser/1997}.

The graph of the energy conditions is given below. 

\begin{figure}
\subfloat[with $k_{4}=-0.43$\label{sfig:testa}]{%
  \includegraphics[scale =0.4]{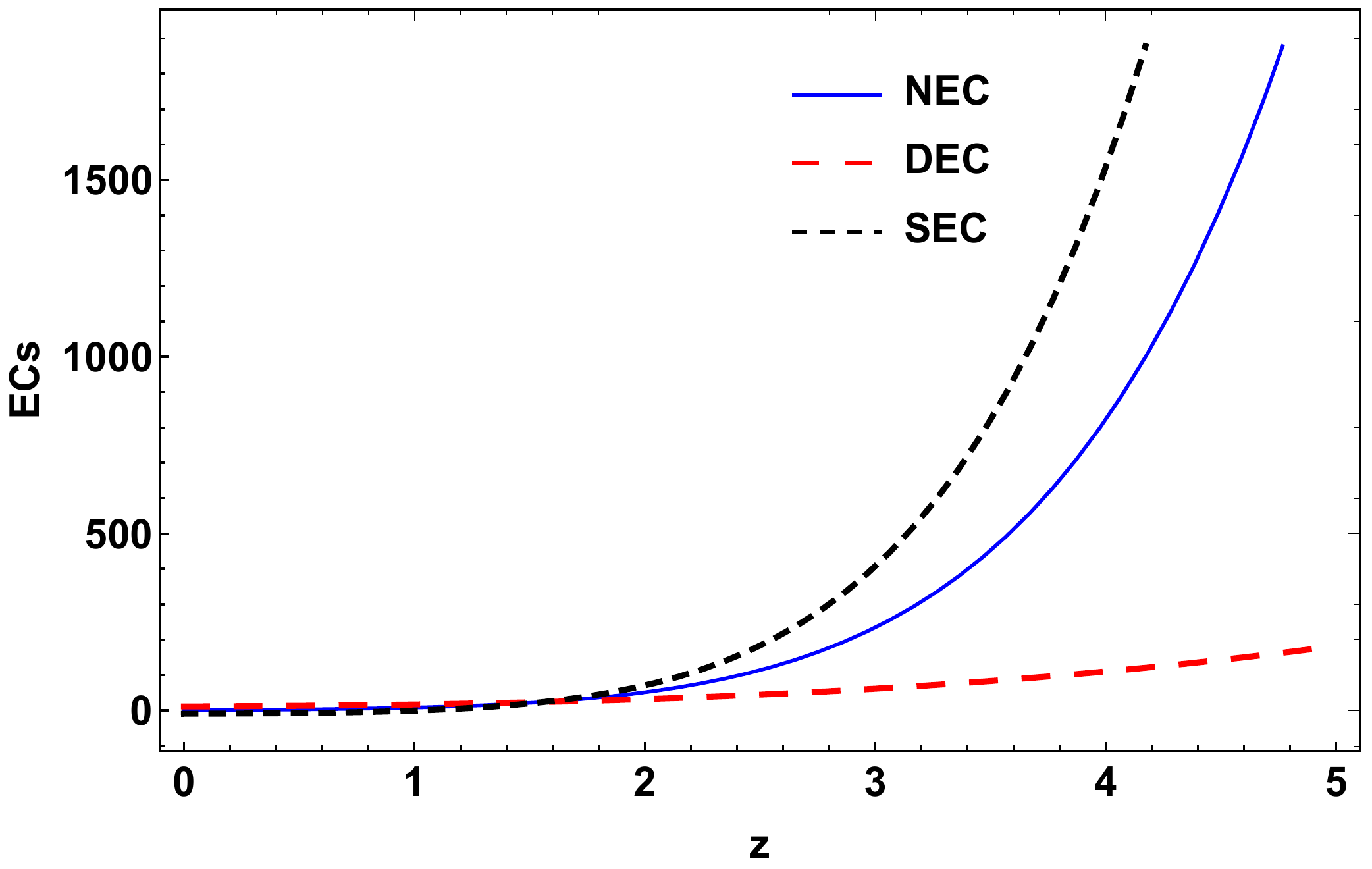}%
}\hfill
\subfloat[with $k_{4}=-0.49$\label{sfig:testa}]{%
  \includegraphics[scale =0.3]{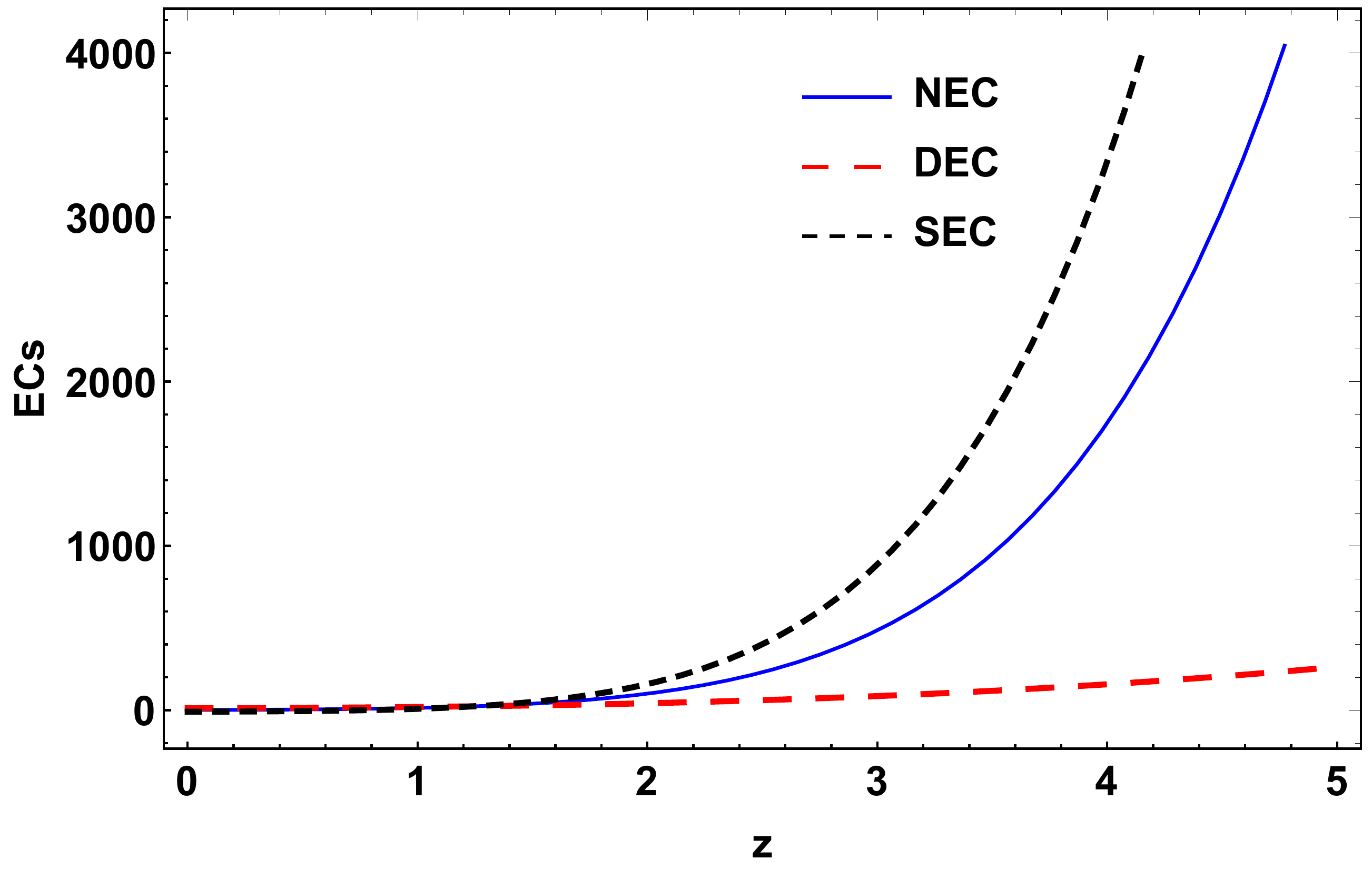}%
}
\caption{Behavior of Energy conditions of the model versus redshift $z$ with 
$\protect\alpha=-0.1$, $\protect\gamma=1.01$, $\protect\omega_{H}=4.1$, $%
\protect\omega_{H2}=1.57$, $\protect\omega_{dH}=-0.1$ and $k_{4}$ as mentioned in (a) \& (b).}
\label{fig5}
\end{figure}

We examine that NEC, DEC hold but SEC violates the model which directly
implies the accelerated expansion of the Universe.

\subsection{Velocity of sound}

The velocity of sound plays a similar role to that of equation of state for
the background cosmology, which relates the pressure and density as in \cite%
{43}, 
\begin{eqnarray*}
c_{s}^{2}= \dfrac{dp}{d\rho}
\end{eqnarray*}
In this study, we have taken speed of light $c$ to be 1, so the stability
condition for the model is $0\leq c_{s}^{2}\leq 1$. The lower bound prevents
dark energy fluctuations from growing exponentially, which can lead to
non-physical situations and the upper one is imposed in order to avoid
super-luminal propagation. Guillermo and Julien \cite{42} reviewed the
concept of sound speed for a cosmological fluid. The non trivial issue of
initial conditions for dark energy perturbations in the radiation era is
studied which is a priori non-adiabatic since $c_{s}^{2} > \omega$. The The
square of the sound speed for bulk viscosity in GR is presented in \cite{29}.

\begin{figure}[]
\centering
\includegraphics[scale =0.3]{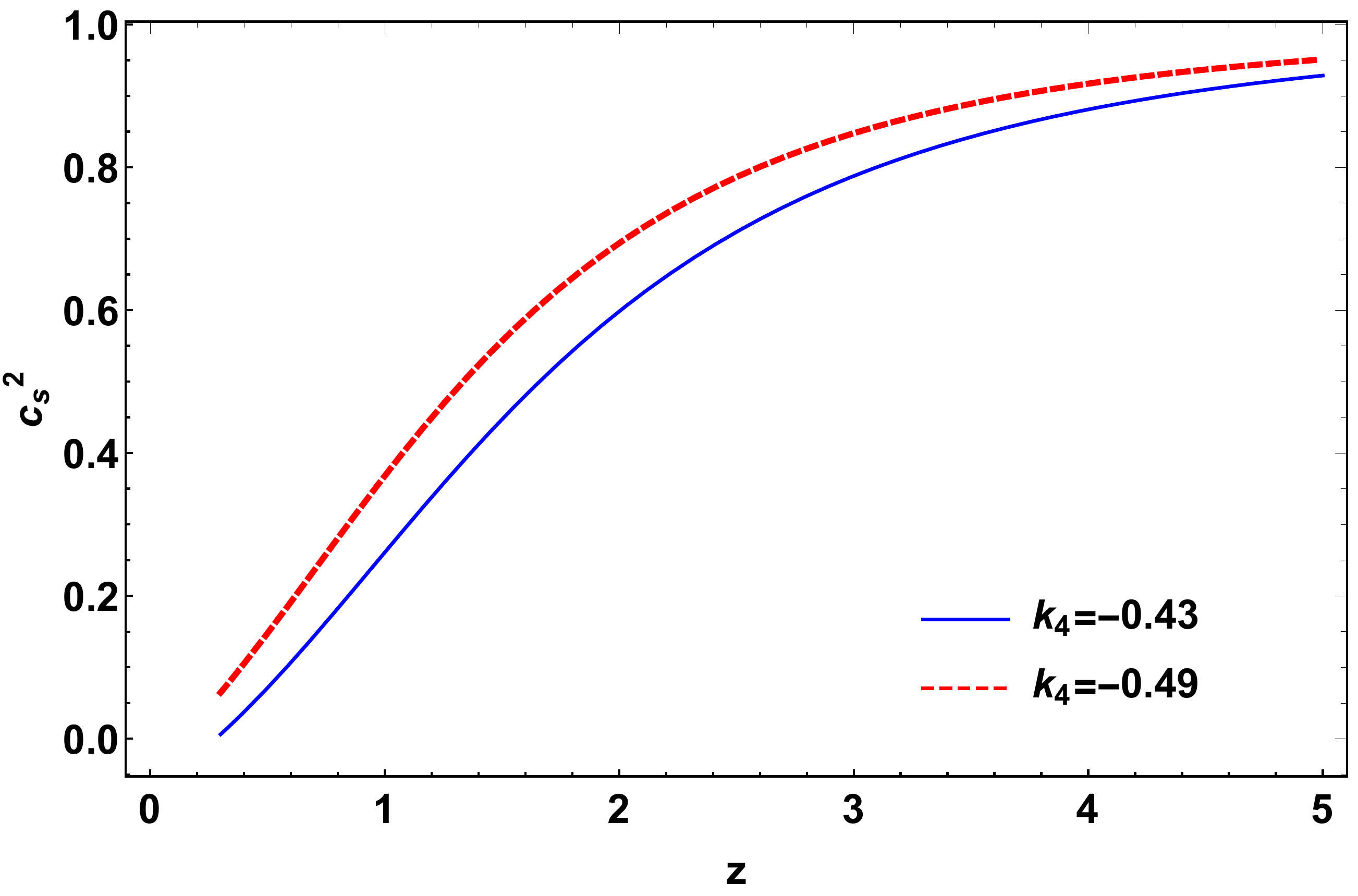}
\caption{Velocity of sound $c_{s}^{2}$ vs redshift z.}
\label{fig6}
\end{figure}

According to the graph above it can be seen that the model satisfies $%
c_{s}^{2}\leq 1$ throughout. Therefore, we can say that our model is stable.

\subsection{Statefinder diagnostics}

In \cite{13}, Sahni et al. have introduced a new cosmological diagnostic
pair \{$r,s$\} - Statefinder. The parameters $r$ and$~s$ are dimensionless
and are constructed from the scale factor $a(t)$ and its time derivatives
similar to the geometrical parameters $H(z)$ and $q(z)$. The Statefinder
help to differentiate and compare between different dark energy models. The
standard cold dark model (SCDM) and the cosmological constant model ($%
\Lambda $CDM) have some fixed points in the $s$-$r$ plane and $q$-$r$ plane.
Any obtained model can be compared with these standard ones to see how a
model approaches or deviates from these models. The expressions $r,s$ for
our model are obtained as, 
\begin{widetext}
\begin{equation}
r= -\frac{\left(\frac{k_{3}^{1/k_{2}}}{k_{4} z+k_{4}}\right)^{-3 k_{2}} \left(\left(\frac{k_{3}^{1/k_{2}}}{k_{4} z+k_{4}}\right)^{k_{2}}-k_{3}\right) \left(-(k_{2}-2) (k_{2}-1) k_{3} \left(\frac{k_{3}^{1/k_{2}}}{k_{4} z+k_{4}}\right)^{k_{2}}+\left(\frac{k_{3}^{1/k_{2}}}{k_{4} z+k_{4}}\right)^{2 k_{2}}+(k_{2}-1) (2 k_{2}-1) k_{3}^2\right)}{\left((k_{4} (z+1))^{k_{2}}-1\right)^3}
\end{equation}
\begin{equation}
s= -\frac{\frac{\left(\left(\frac{k_{3}^{1/k_{2}}}{k_{4} z+k_{4}}\right)^{k_{2}}-k_{3}\right) \left(-(k_{2}-2) (k_{2}-1) k_{3} \left(\frac{k_{3}^{1/k_{2}}}{k_{4} z+k_{4}}\right)^{k_{2}}+\left(\frac{k_{3}^{1/k_{2}}}{k_{4} z+k_{4}}\right)^{2 k_{2}}+(k_{2}-1) (2 k_{2}-1) k_{3}^2\right) \left(\frac{k_{3}^{1/k_{2}}}{k_{4} z+k_{4}}\right)^{-3 k_{2}}}{\left((k_{4} (z+1))^{k_{2}}-1\right)^3}+1}{3 \left(\frac{k_{2}}{(k_{4} (z+1))^{k_{2}}-1}+k_{2}-1.5\right)}
\end{equation}
\end{widetext}

Although, the presented model do not contain any extra source term (dark
energy) but the bulk viscous term exerts extra pressure and plays the role
of the dark energy. The following are plots show the behavior of our
obtained model compared with the SCDM and $\Lambda $CDM models.

\begin{figure}[]
\centering
\includegraphics[scale =0.4]{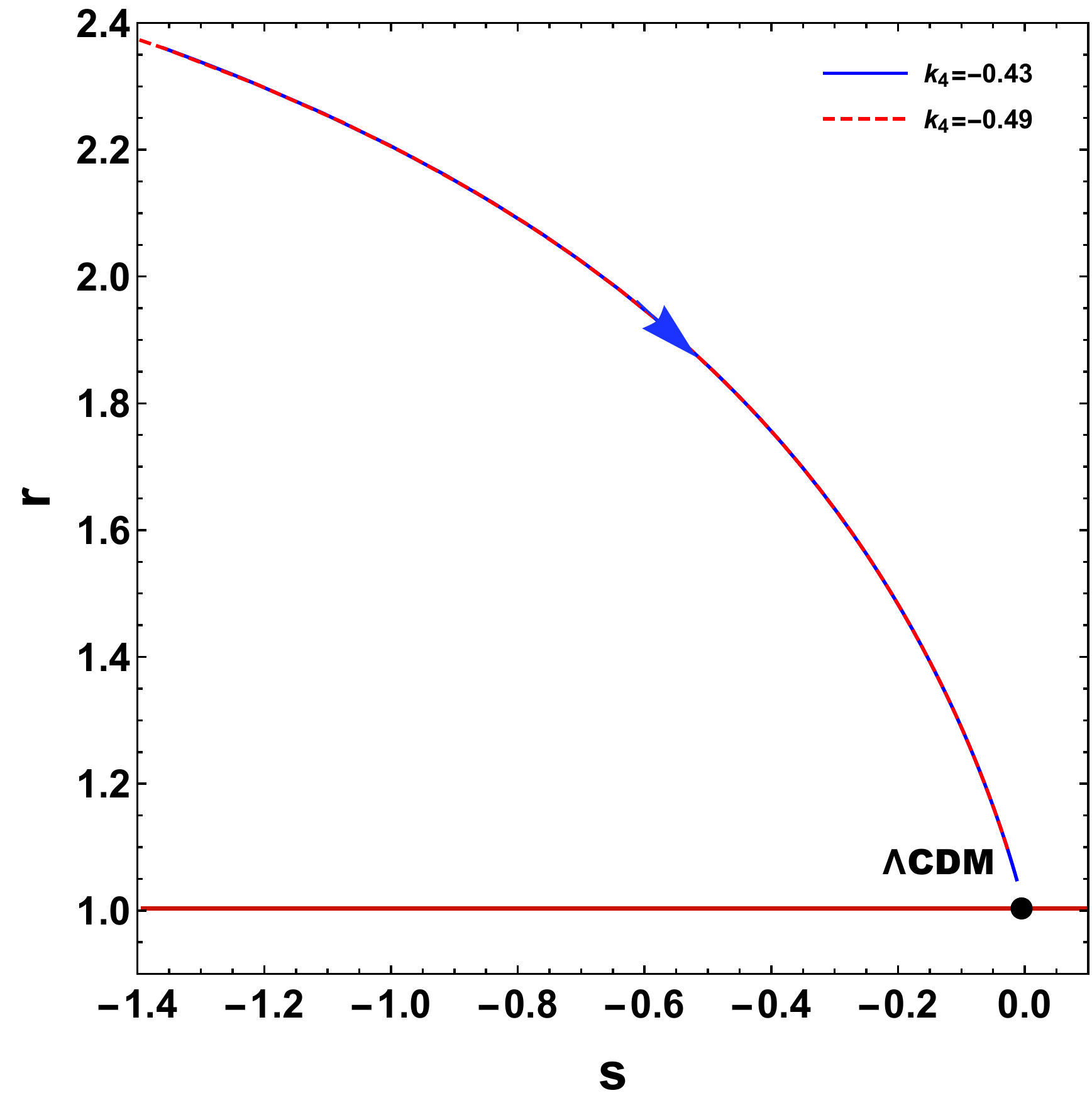}
\caption{The plot shows the behavior of the presented model in the $s$-$r$
plane and $\Lambda $CDM model.}
\label{fig7}
\end{figure}

\begin{figure}[]
\centering
\includegraphics[scale =0.4]{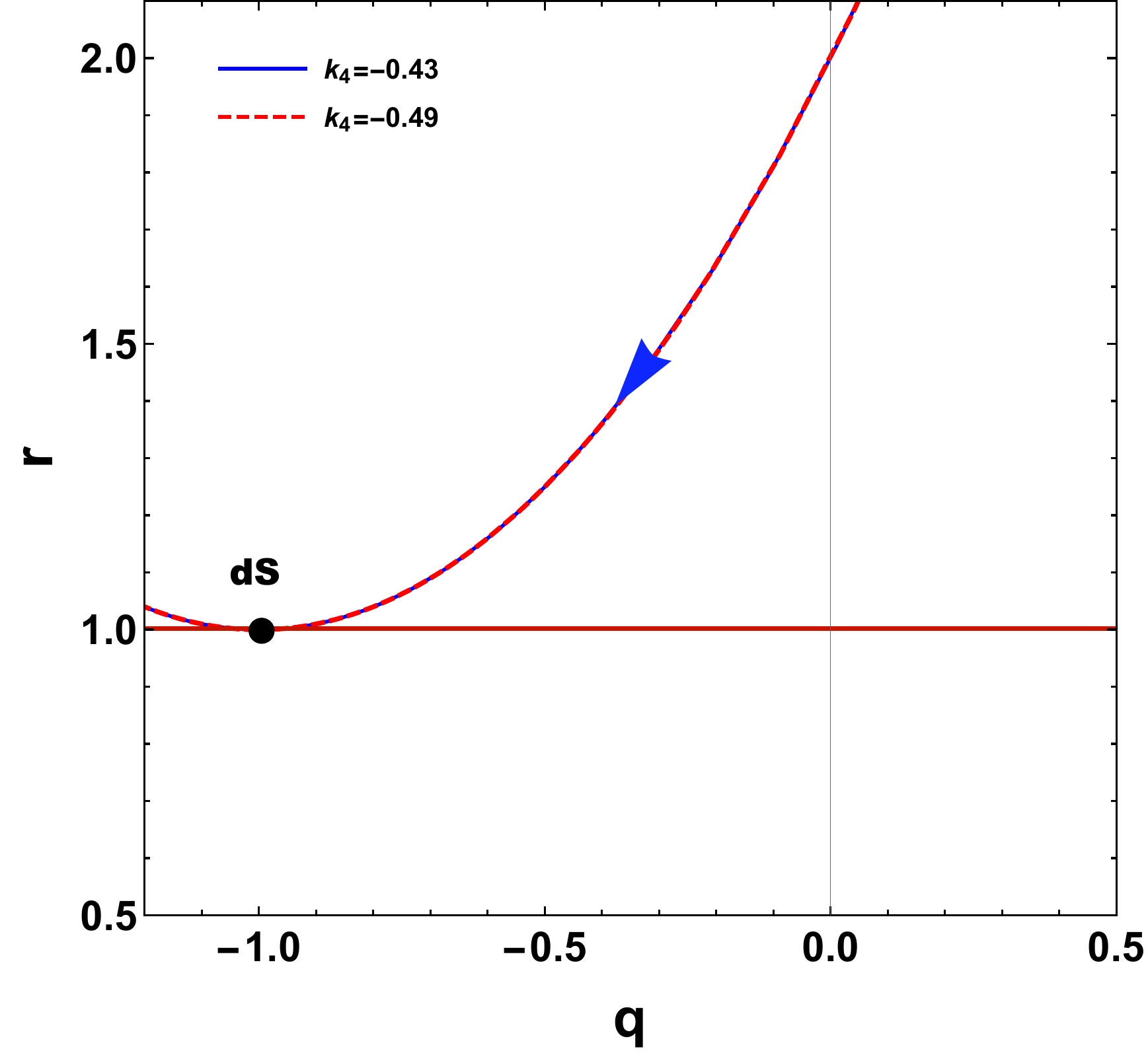}
\caption{The plot shows the behavior of the presented model in the $q$-$r$
plane $\Lambda $CDM model.}
\label{fig8}
\end{figure}

In the Fig.\ref{fig7}, the model behavior is shown in ${s}${-}${r}$ plane
which is somewhat similar to \cite{Titus}. The point $(s,r)=(0,1)$
corresponds to the $\Lambda $CDM of the universe. Our model is also
resembling to the $\Lambda $CDM in future and ultimately freezing to it.
Similarly, Fig.\ref{fig8} shows that our model approaching to the de Sitter
point ($q=-1,r=1$) and deviated from the SCDM model.

\subsection{$Om$ diagnostic}

Now, we turn to a discussion on $Om$ diagnostic written as $Om(z)$. $Om(z)$
is used to differentiate standard $\Lambda $CDM model from various dark
energy models \cite{15}. In the analysis of $Om$ diagnostic only first order
derivative are used as it involves the Hubble parameter depending on a
single time derivative of $a(t)$. In reference with Sahni et al. \cite{16}
and Zunckel and Clarkson \cite{17}, $Om(z)$ for flat universe is defined as 
\newline
\begin{equation}
Om(z)=\frac{\left( \frac{H(z)}{H_{0}}\right) ^{2}-1}{(1+z)^{3}-1}
\end{equation}

\begin{figure}[]
\centering
\includegraphics[scale =0.35]{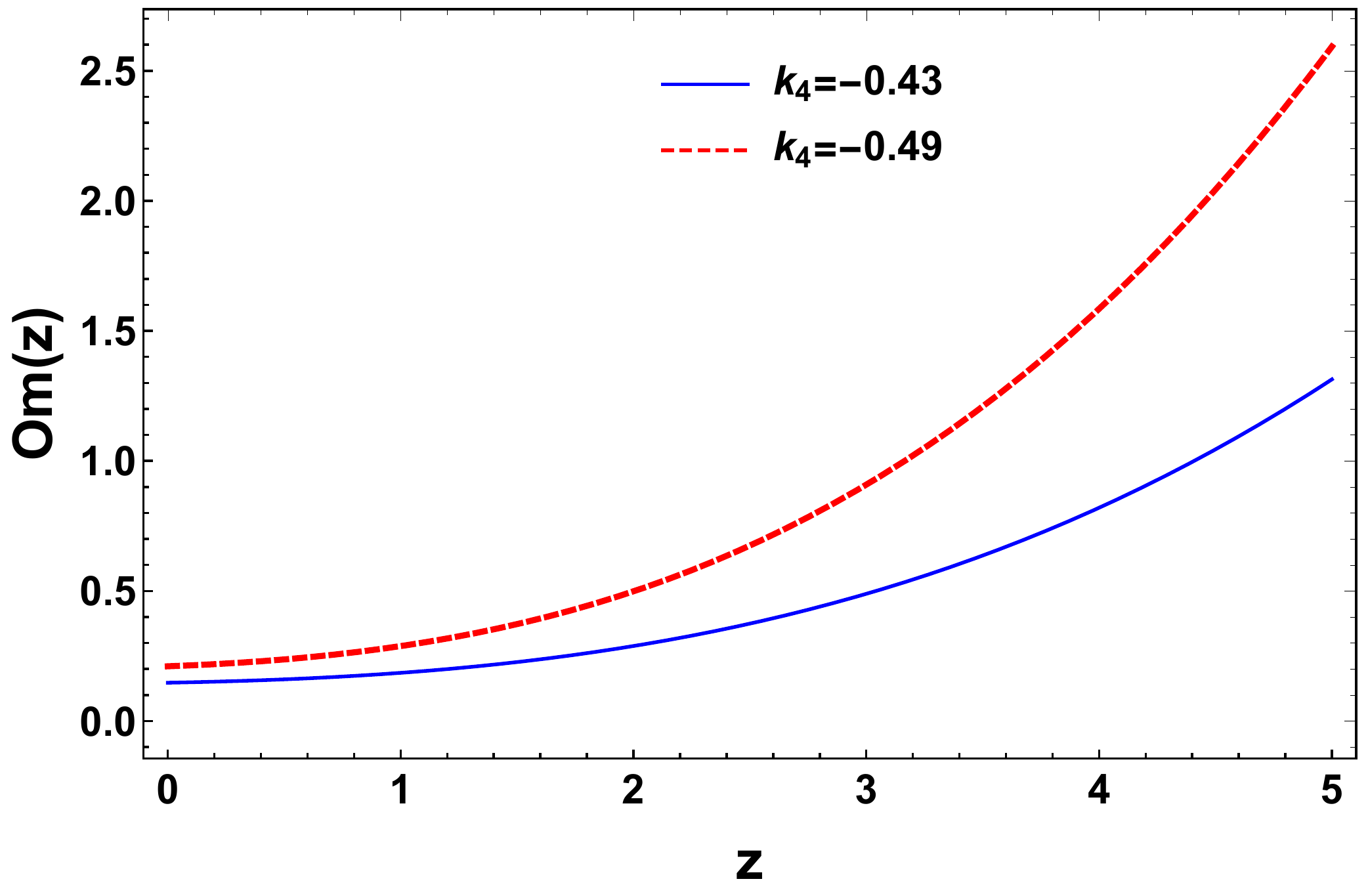}
\caption{Behavior of $Om$ versus redshift $z$}
\label{fig9}
\end{figure}

Thus, we have different values of $Om(z)$ for the $\Lambda $CDM model,
phantom and quintessence cosmological models. According to the curvature
variation, we can describe the behavior of dark energy as quintessence type (%
$\omega >-1$) corresponding to negative curvature, phantom type ($\omega <-1$%
) corresponding to its positive curvature and $Om(z)$=$\Lambda $CDM to zero
curvature. The parametrization of $Om(z)$ is done in \cite{41} to show how
the combination of most recent and naturally improved observations about the 
$H(z)$ and SNeIa be implemented to study the consistency or acknowledge the
tension between the $\Lambda $CDM model and observations. The behavior can
be easily seen in Fig. \ref{fig9} which shows that at late times, the growth
of $Om(z)$ favors the decaying dark energy models as discussed in \cite{18}. 
\begin{equation}
Om(z)=\frac{\frac{1.\left( (k_{4}z+k_{4})^{k_{2}}-1\right) ^{2}}{\left(
k_{4}^{k_{2}}-1\right) ^{2}}-1}{(z+1)^{3}-1}
\end{equation}

\subsection{Fitting the model with $H(z)$ \& $SNIa$ datasets}

Study of the structure, the origin and the evolution of the universe through
observations is known as observational cosmology. Several types of
observational datasets are available at present for different measurements
such as Type Ia Supernovae \cite{1,2} data, Cosmic Microwave Background
Radiation \cite{30} data, Baryon Acoustic Oscillations \cite{31} data,
Planck data etc. and are some spectacular observations providing strong
evidence for the acceleration of the universe. So, we shall check the
viability of our obtained model with any of these datasets. Here, we have
taken into account $57$ points of $H(z)$ data (Appendix Table 2.), wherein $%
31$ points of Hubble data points are from the differential age method and $26
$ points are from BAO and other methods \cite{sharov}. Secondly, we have
taken into account $580$ points of type Ia Supernovae from Union $2.1$
compilation datasets \cite{Union2.1 DATA, ritika/2018} to achieve our goal
to find best fit values of the model parameters and compare to the $\Lambda $%
CDM model.

The $\chi ^{2}$ function for the $H(z)$ datasets is taken to be 
\begin{equation}
\chi _{H}^{2}=\sum_{i=1}^{57}\dfrac{\left[ H^{obs}(z_{i})-H^{th}(z_{i})%
\right] ^{2}}{\sigma (z_{i})^{2}}
\end{equation}%
where $H^{obs}$ and $H^{th}$ are the observed and theoretical value of $H$
and also $\sigma (z_{i})$ is the standard error in the measured value of $H$%
. The following plot shows nice fit to the $H(z)$ datasets with suitable
model parameter values compared with $\Lambda $CDM model. We have taken $%
H_{0}=67.66$ $km/sec/Mpc$ for our calculation.

\begin{figure}[]
\centering
\includegraphics[scale =0.5]{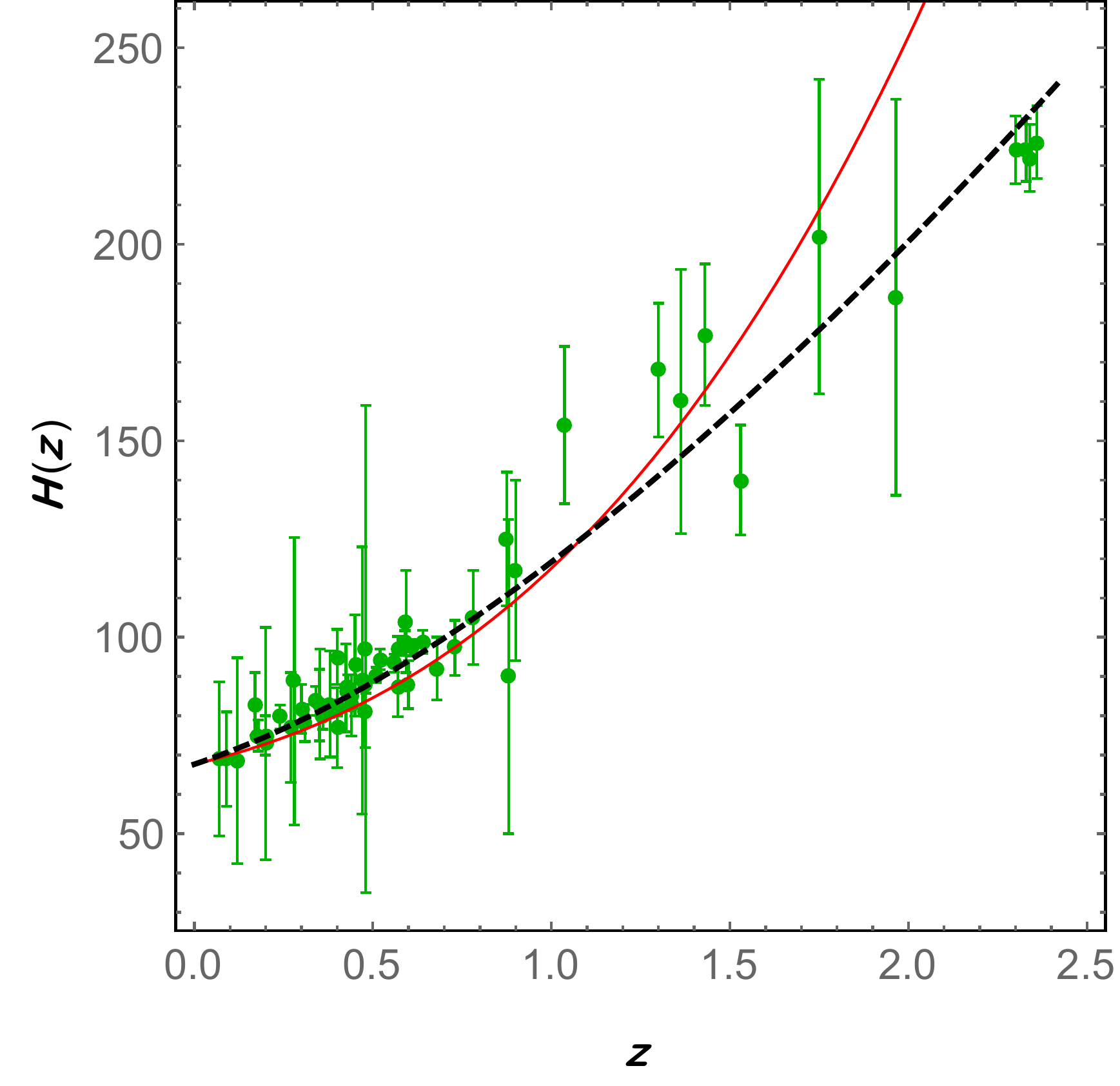}
\caption{The plot shows the $57$ points of $H(z)$ datasets (Blue dots) with
corresponding error bars along with the presented model (solid red line)
which has a better fit to the $H(z)$ datasets for $k_{2}=3$ and $k_{4}=-0.49$%
. $\Lambda $CDM model is also shown in black dashed line for model
comparision. }
\label{fig10}
\end{figure}

The $\chi ^{2}$ function for the type Ia supernovae datasets is taken to be 
\begin{equation}
\chi _{SN}^{2}=\sum_{i=1}^{580}\frac{\left[ \mu _{th}(\mu _{0},z_{i})-\mu
_{obs}(z_{i})\right] ^{2}}{\sigma _{\mu (z_{i})}^{2}},
\end{equation}%
where $\mu _{obs}$, $\mu _{th}$, $\sigma _{\mu (z_{i})}$, denotes the
observed and theoretical distance modulus of the model, the standard error
in the measurement of $\mu (z)$ respectively. We fit free parameters of our
model, comparing $\mu _{obs}$ with theoretical values $\mu _{th}$ of
distance modulus. The distance model $\mu (z)$ is given by 
\begin{equation}
\mu (z)=m=m^{\prime }+5LogD_{l}(z)+\mu _{0},
\end{equation}%
where $D_{l}(z)$ and $\mu _{0}$ are the luminosity distance and nuisance
parameter respectively. Also $m$ and $m^{\prime }$ serve as the apparent and
absolute magnitudes of standard candle respectively. We calculate the $%
\chi_{SN}^{2}$ function and the distance $D_{l}(z)$ that measures
differences between the $SNeIa$ observational data and predictions of a
model. The following plot shows nice fit to the $SNIa$ datasets with
suitable model parameter values compared with $\Lambda $CDM model.

\begin{figure}[tbp]
\centering
\includegraphics[scale =0.5]{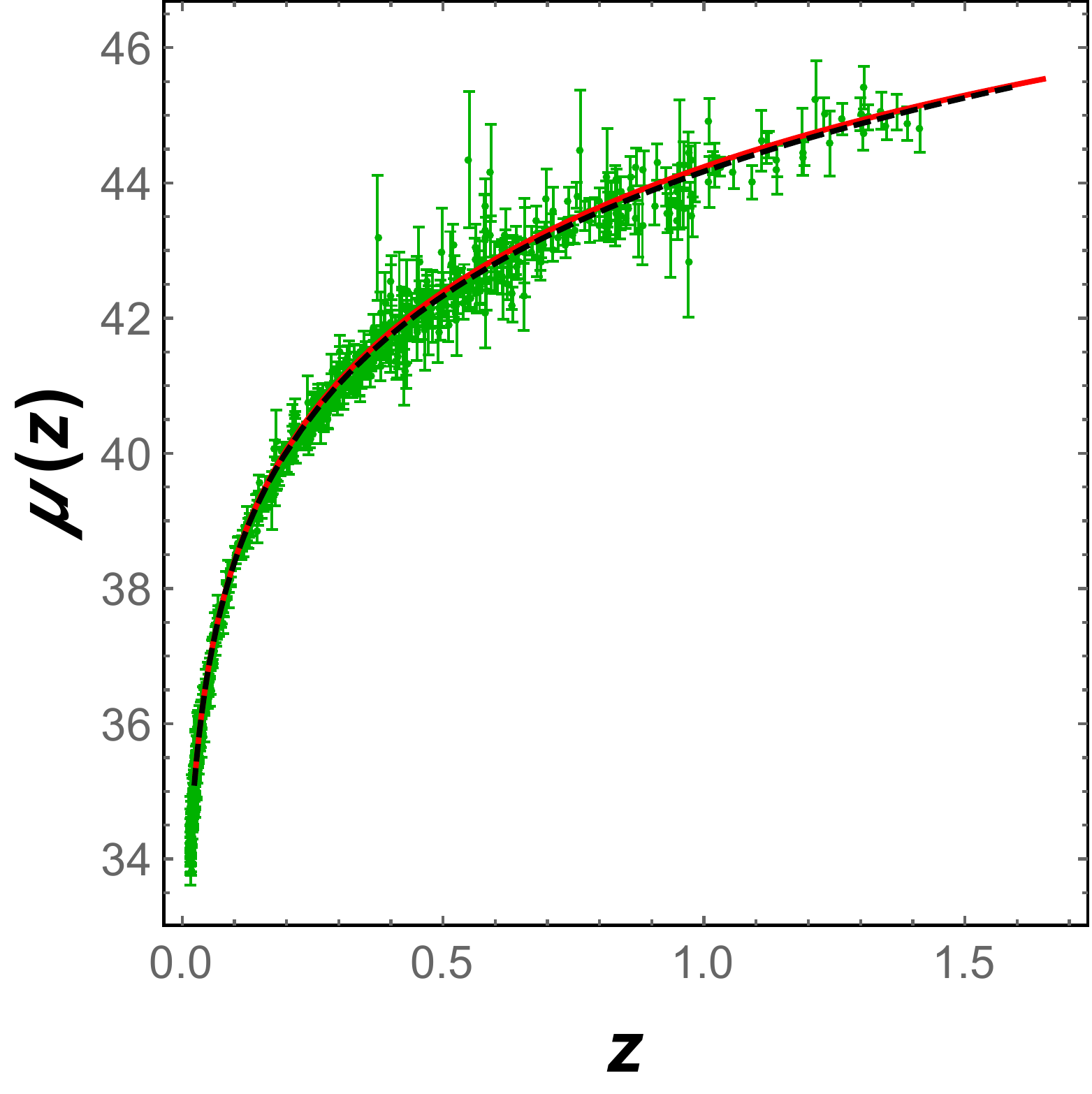}
\caption{The plot shows the $580$ points of $SNIa$ datasets (Blue dots) with
corresponding error bars along with the presented model (solid red line)
which has a better fit to the $SNIa$ datasets for $k_{2}=3$ and $k_{4}=-0.49$%
. $\Lambda $CDM model is also shown in black dashed line for model
comparision.\ \ \ \ \ \ \ \ \ \ \ \ \ \ \ \ \ \ \ \ \ \ \ \ \ \ \ \ \ \ \ \
\ }
\label{fig11}
\end{figure}

\subsection{Estimation of model parameters with $H(z)$, $SNIa$ \& $BAO$
datasets}\label{F}

We can see, in the expression Eq.\eqref{e21}, we have only two model
parameters $k_{2}$ and $k_{4}$. Here, in this subsection, we shall find the
constraints with the above discussed datasets i.e. $H(z)$ and $SNIa$
together with one more external data, the Baryon Acoustic Oscillation (BAO)
datasets for our analysis. The chi square value corresponding to BAO
measurements is given by \cite{gio} 
\begin{equation}
\chi _{BAO}^{2}=B^{T}C^{-1}B,
\end{equation}%
where the matrices $B$, inverse covariance matrix $C^{-1}$ and the data
details are discussed in the appendix.

With these three samples of datasets, we have found the likelihood contours
for the model parameters $k_{2}$ and $k_{4}$ at $1$-$\sigma $, $2$-$\sigma $
and $3$-$\sigma $ level and are plotted in the $k_{2}$-$k_{4}$ plane as
shown in the figures. We have found constraints with independent $H(z)$
datasets and combined $Hz+SNIa+BAO$ datasets. The best estimated values of
the model parameters $k_{2}$ and $k_{4}$ are found to be $k_{2}=3$, $%
k_{4}=-0.4389$ and $k_{2}=3$, $k_{4}=-0.43374$ respectively for independent $%
H(z)$ datasets and joint $Hz+SNIa+BAO$ datasets. 
\begin{figure}[]
\centering
\includegraphics[scale =0.4]{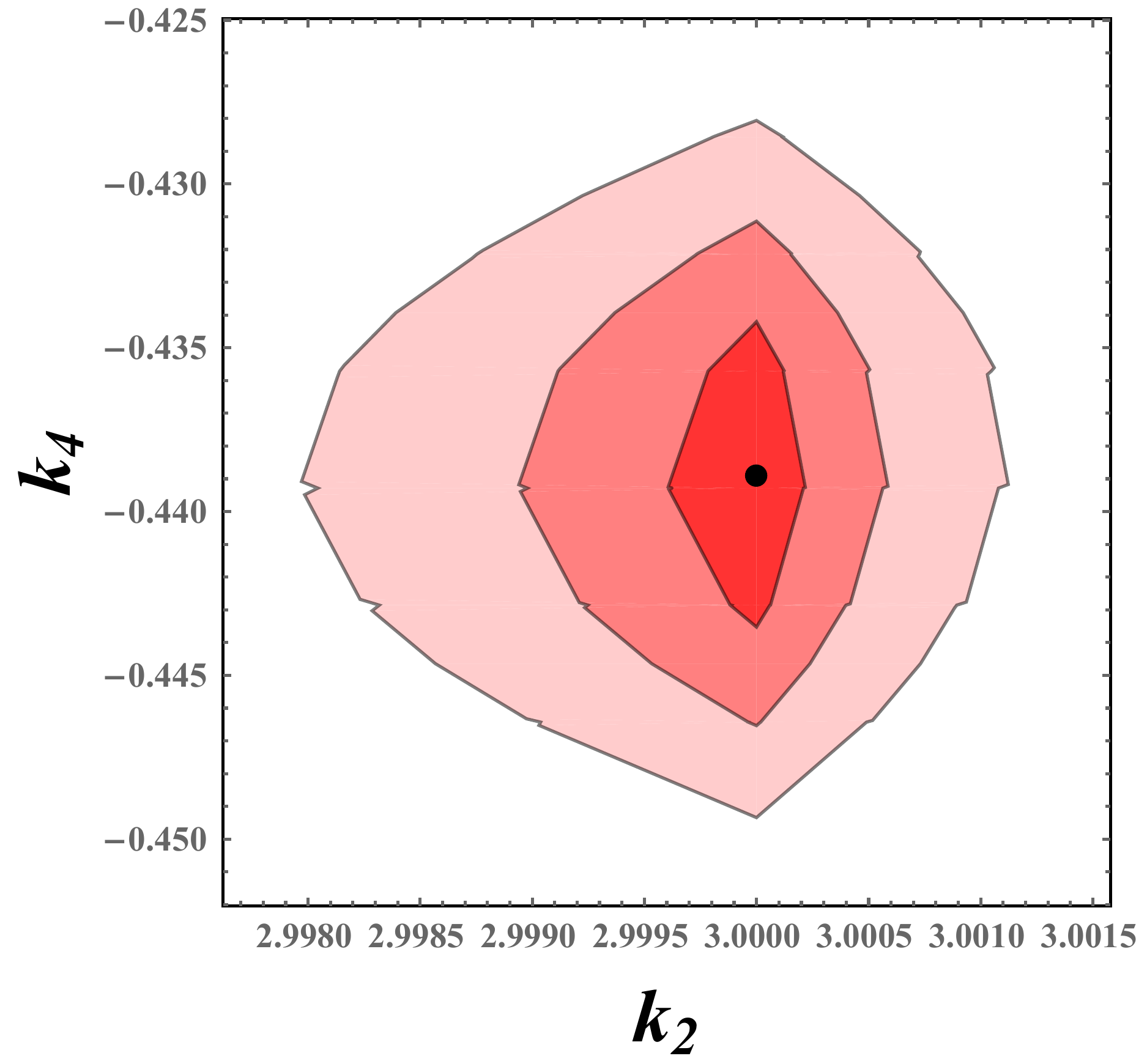}
\caption{The plot shows the contour plot for the model parameters $k_{2}$
and $k_{4}$ for independent $H(z)$ datasets at $1$-$\protect\sigma $, $2$-$%
\protect\sigma $ and $3$-$\protect\sigma $ level in $k_{2}$-$k_{4}$ plane.
The best estimated value is $k_{2}=3$ and $k_{4}=-0.4389$.}
\label{fig12}
\end{figure}

\begin{figure}[]
\centering
\includegraphics[scale =0.5]{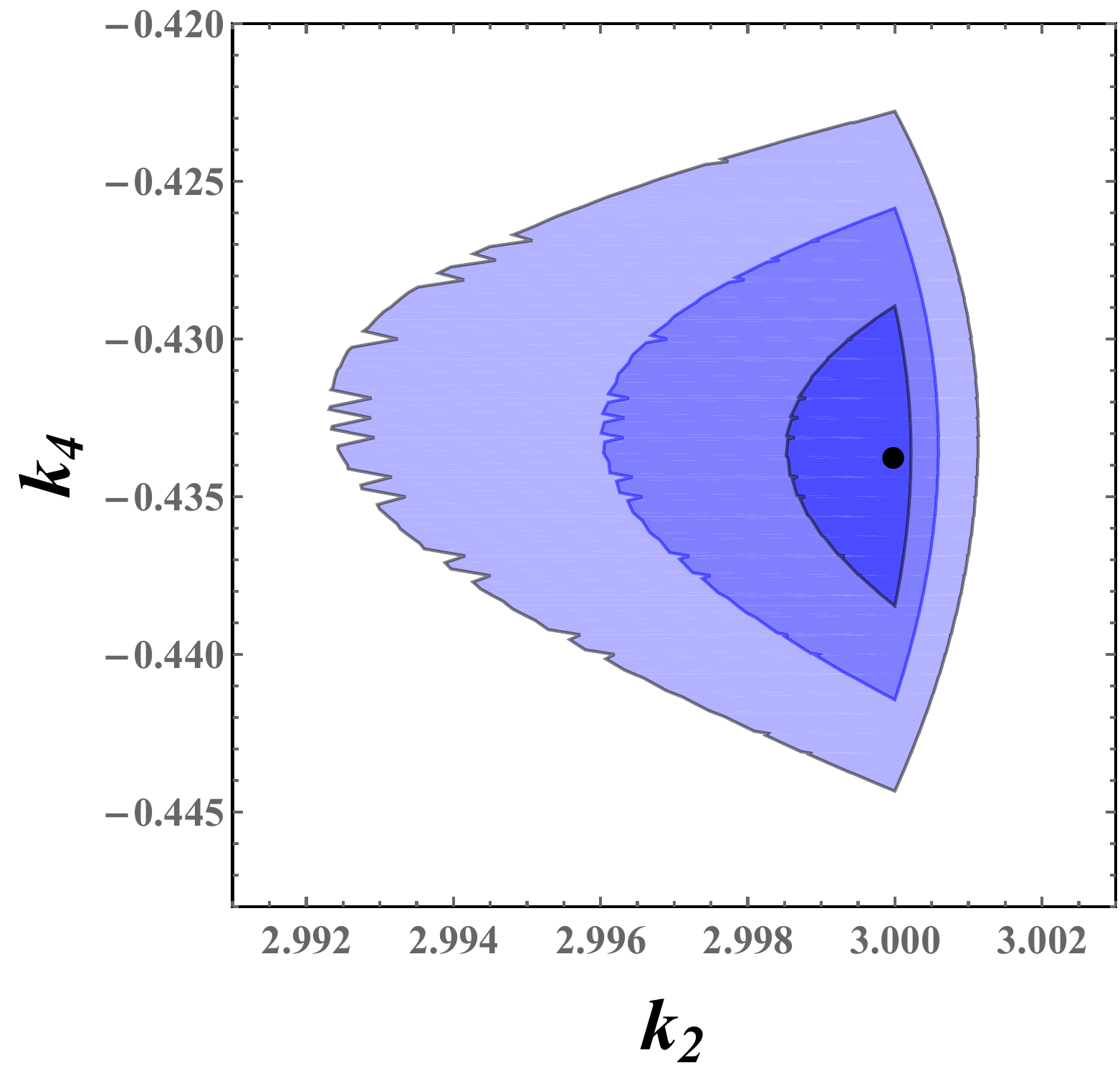}
\caption{The plot shows the contour plot for the model parameters $k_{2}$
and $k_{4}$ for combined $Hz+SNIa+BAO$ datasets at $1$-$\protect\sigma $, $2$%
-$\protect\sigma $ and $3$-$\protect\sigma $ level in $k_{2}$-$k_{4}$
plane.The best estimated value is $k_{2}=3$ and $k_{4}=-0.43374$.}
\label{fig13}
\end{figure}

\section{Conclusion}\label{V}

In this article, we have studied a cosmological model in which we have
discussed the phenomenon of cosmic acceleration without the need of dark
energy but with a viscous fluid. With this effective viscosity EoS, the
dynamical equation of the Hubble parameter is completely integrable and an
exact solution for Einstein's field equation is obtained in modified $f(R,T)$
gravity in the FLRW background. The effective EoS (pressure with additional
bulk viscosity) describes the late-time acceleration of the Universe without
introducing a cosmological constant or dark energy. We show that the matter
described by an effective viscosity EoS can fit the observational data well,
so the present effective viscosity model may be considered an alternative
candidate to explain the late-time accelerating expansion of the Universe.

The deceleration parameter shows a signature flip from early deceleration to
present acceleration at $z_{t}\approx 0.845815$ and $z_{t}\approx 0.619797$
for $k_{4}=-0.43$ and $k_{4}=-0.49$ respectively with a negative value of
the $q_{0}\approx -0.779046$ and $q_{0}\approx -0.684206$, see Fig. \ref%
{fig1}. The evolution of the effective equation of state parameter $\omega $
is shown in Fig. \ref{fig4} showing the negative value of the $\omega
_{0}\approx -0.888046$ for $k_{4}=-0.43$ and $\omega _{0}\approx -0.838394$
for $k_{4}=-0.49$ remains in the quintessence region (do\ not cross the
phantom divide line $\omega =-1$) approaching to $-1$ in the infinite future
leading to Einstein-de-Sitter model. It is also worth mentioning that the
EoS parameter $\omega $ shows a transition from positive pressure regime in
the past to a negative pressure regime at present era implying that the
incorporation of bulk viscous pressure term in the model plays a vital role
for rendering a decelerating expansion in the past (suitable for structure
formation) and an accelerated expansion at present. This can be seen in Fig. %
\ref{fig4}, which confirms from the standard cosmology that the latter
regime may happen when $\omega <-\frac{1}{3}$. In the case, when $\alpha =0$
($\alpha $ is the coupling constant for modified gravity), the field
equations \eqref{e8} and \eqref{e9} will reduce to general relativity and we
can't get the same conditions where the pressure becomes negative throughout
the evolution and the plot for $w(z)$ remains in negative part (not shown).
So, we can say that the coupling constant of modified gravity play a major
role in this context.

We have discussed some physical characteristics of the model and discussed
the evolution of physical parameters together with the Energy Conditions. It
is seen that NEC, DEC does not violate the model but SEC fails to satisfy,
which produces a repulsive force and make the Universe to get jerk. The
violation of SEC in Fig. \ref{fig5} shows the viability of our model as
mentioned in \cite{Barcelo/2002}. We have also discussed the velocity of
sound favoring our model's consistency. Moreover, analysis of statefinder
parameters and $Om$ diagnostic also have been done and compared with the $%
\Lambda $CDM model. Finally, we have fitted our model with the updated $57$
points of Hubble datasets and $580$ points of Union $2.1$ compilation
Supernovae datasets compared with the $\Lambda$CDM model.

ECs have provided us with special insights into the deep structure for space
and time in the cosmic space-time evolution processes. In the present model
NEC and DEC validated whereas SEC is violated as per the requirement of
cosmic acceleration (see Fig. \ref{fig5}). As we know the wormhole formation
requires explicitly the null energy condition violation, which attracts us
to further study the dark energy confrontation with the uncommon space time
structure. We will publish the related work elsewhere soon. Further study
can be done with this effective viscosity EoS in non-minimally coupled
gravity.

\acknowledgements SA acknowledges CSIR, New Delhi, India for JRF. PKS
acknowledges CSIR, New Delhi, India for financial support to carry out the
Research project [No.03(1454)/19/EMR-II Dt.02/08/2019]. We are very much
grateful to the honorable referees and the editor for illuminating
suggestions that have significantly improved our work in terms of research
quality as well as the presentation.

\section*{Appendix}

Details of $H(z)$ datasets: The $57$ points of Hubble parameter values $H(z)$
with errors $\sigma _{H}$ from differential age ($31$ points) method and BAO
and other ($26$ points) methods are shown in the following table.
\begin{widetext}
\begin{center}
\begin{tabular}{|c|c|c|c|c|c|c|c|}
\hline
$z$ & $H(z)$ & $\sigma _{H}$ & Ref. & $z$ & $H(z)$ & $\sigma _{H}$ & Ref. \\ 
\hline
$0.070$ & $69$ & $19.6$ & \cite{h14} & $0.24$ & $79.69$ & $2.99$ & \cite{h7}
\\ \hline
$0.90$ & $69$ & $12$ & \cite{h13} & $0.30$ & $81.7$ & $6.22$ & \cite{h10} \\ 
\hline
$0.120$ & $68.6$ & $26.2$ & \cite{h14} & $0.31$ & $78.18$ & $4.74$ & \cite%
{h6} \\ \hline
$0.170$ & $83$ & $8$ & \cite{h13} & $0.34$ & $83.8$ & $3.66$ & \cite{h7} \\ 
\hline
$0.1791$ & $75$ & $4$ & \cite{h16} & $0.35$ & $82.7$ & $9.1$ & \cite{h1} \\ 
\hline
$0.1993$ & $75$ & $5$ & \cite{h16} & $0.36$ & $79.94$ & $3.38$ & \cite{h6}
\\ \hline
$0.200$ & $72.9$ & $29.6$ & \cite{h15} & $0.38$ & $81.5$ & $1.9$ & \cite{h11}
\\ \hline
$0.270$ & $77$ & $14$ & \cite{h13} & $0.40$ & $82.04$ & $2.03$ & \cite{h6}
\\ \hline
$0.280$ & $88.8$ & $36.6$ & \cite{h15} & $0.43$ & $86.45$ & $3.97$ & \cite%
{h7} \\ \hline
$0.3519$ & $83$ & $14$ & \cite{h16} & $0.44$ & $82.6$ & $7.8$ & \cite{h8} \\ 
\hline
$0.3802$ & $83$ & $13.5$ & \cite{h18} & $0.44$ & $84.81$ & $1.83$ & \cite{h6}
\\ \hline
$0.400$ & $95$ & $17$ & \cite{h13} & $0.48$ & $87.79$ & $2.03$ & \cite{h6}
\\ \hline
$0.4004$ & $77$ & $10.2$ & \cite{h18} & $0.51$ & $90.4$ & $1.9$ & \cite{h11}
\\ \hline
$0.4247$ & $87.1$ & $11.2$ & \cite{h18} & $0.52$ & $94.35$ & $2.64$ & \cite%
{h6} \\ \hline
$0.4497$ & $92.8$ & $12.9$ & \cite{h18} & $0.56$ & $93.34$ & $2.3$ & \cite%
{h6} \\ \hline
$0.470$ & $89$ & $34$ & \cite{h19} & $0.57$ & $87.6$ & $7.8$ & \cite{h2} \\ 
\hline
$0.4783$ & $80.9$ & $9$ & \cite{h18} & $0.57$ & $96.8$ & $3.4$ & \cite{h5}
\\ \hline
$0.480$ & $97$ & $62$ & \cite{h14} & $0.59$ & $98.48$ & $3.18$ & \cite{h6}
\\ \hline
$0.593$ & $104$ & $13$ & \cite{h16} & $0.60$ & $87.9$ & $6.1$ & \cite{h8} \\ 
\hline
$0.6797$ & $92$ & $8$ & \cite{h16} & $0.61$ & $97.3$ & $2.1$ & \cite{h11} \\ 
\hline
$0.7812$ & $105$ & $12$ & \cite{h16} & $0.64$ & $98.82$ & $2.98$ & \cite{h6}
\\ \hline
$0.8754$ & $125$ & $17$ & \cite{h16} & $0.73$ & $97.3$ & $7.0$ & \cite{h8}
\\ \hline
$0.880$ & $90$ & $40$ & \cite{h14} & $2.30$ & $224$ & $8.6$ & \cite{h9} \\ 
\hline
$0.900$ & $117$ & $23$ & \cite{h13} & $2.33$ & $224$ & $8$ & \cite{h12} \\ 
\hline
$1.037$ & $154$ & $20$ & \cite{h16} & $2.34$ & $222$ & $8.5$ & \cite{h4} \\ 
\hline
$1.300$ & $168$ & $17$ & \cite{h13} & $2.36$ & $226$ & $9.3$ & \cite{h3} \\ 
\hline
$1.363$ & $160$ & $33.6$ & \cite{h17} &  &  &  &  \\ \hline
$1.430$ & $177$ & $18$ & \cite{h13} &  &  &  &  \\ \hline
$1.530$ & $140$ & $14$ & \cite{h13} &  &  &  &  \\ \hline
$1.750$ & $202$ & $40$ & \cite{h13} &  &  &  &  \\ \hline
$1.965$ & $186.5$ & $50.4$ & \cite{h17} &  &  &  &  \\ \hline
\end{tabular}
\end{center}
\end{widetext}

Details of BAO datasets: From very large scales, Baryon Acoustic Oscillation
measures the structures in the universe. In this article, we have considered
the sample of BAO distances measurements from surveys of SDSS(R) \cite{padn}%
, 6dF Galaxy survey \cite{6df}, BOSS CMASS \cite{boss} and WiggleZ \cite{wig}%
. So, the distance redshift ratio $d_{z}$ is given as $d_{z}=\frac{%
r_{s}(z_{\ast })}{D_{v}(z)},$where $r_{s}(z_{\ast })$ is the co-moving sound
horizon at the time photons decouple and $z_{\ast }$ is the photon
decoupling redshift. In accordance to Planck 2015 results \cite{Hz-Plank}, $%
z_{\ast }=1090$. We have taken $r_{s}(z_{\ast })$ as considered in \cite%
{waga}. Also, dilation scale is read as $D_{v}(z)$ and is given by $D_{v}(z)=%
\big(\frac{d_{B}^{2}(z)z}{H(z)}\big)^{\frac{1}{3}}$, where $d_{A}(z)$ is the
angular diameter distance. The matrix $B$ in the chi square formula of BAO
datasets is given by $d_{B}(z_{\ast })/D_{V}(z_{BAO})$ and is calculated as 
\begin{equation*}
B=\left( 
\begin{array}{c}
\frac{d_{B}(z_{\star })}{D_{V}(0.106)}-30.95 \\ 
\frac{d_{B}(z_{\star })}{D_{V}(0.2)}-17.55 \\ 
\frac{d_{B}(z_{\star })}{D_{V}(0.35)}-10.11 \\ 
\frac{d_{B}(z_{\star })}{D_{V}(0.44)}-8.44 \\ 
\frac{d_{B}(z_{\star })}{D_{V}(0.6)}-6.69 \\ 
\frac{d_{B}(z_{\star })}{D_{V}(0.73)}-5.45%
\end{array}%
\right) \,,
\end{equation*}%
and the inverse covariance matrix $C^{-1}$ defined in \cite{gio} is given
by 
\begin{widetext}
\begin{equation*}
C^{-1}=\left( 
\begin{array}{cccccc}
0.48435 & -0.101383 & -0.164945 & -0.0305703 & -0.097874 & -0.106738 \\ 
-0.101383 & 3.2882 & -2.45497 & -0.0787898 & -0.252254 & -0.2751 \\ 
-0.164945 & -2.454987 & 9.55916 & -0.128187 & -0.410404 & -0.447574 \\ 
-0.0305703 & -0.0787898 & -0.128187 & 2.78728 & -2.75632 & 1.16437 \\ 
-0.097874 & -0.252254 & -0.410404 & -2.75632 & 14.9245 & -7.32441 \\ 
-0.106738 & -0.2751 & -0.447574 & 1.16437 & -7.32441 & 14.5022%
\end{array}%
\right) \,.
\end{equation*}
\end{widetext}.


\begin{thebibliography}{99}
\bibitem{1} A. G. Riess et al., Astron. J. \textbf{116}, 1009 (1998).

\bibitem{2} S. Perlmutter et al., Astrophys. J. \textbf{517}, 565 (1999).

\bibitem{3} P. M. Garnavich et al., Astrphys. J. \textbf{493}, L53 (1998).

\bibitem{4} S. Nojiri, S.D. Odintsov, Phys. Rep. \textbf{505}, 59 (2011).

\bibitem{5} K. Bamba, S.D Odintsov, Symmetry \textbf{7}, 220 (2015).

\bibitem{Yousaf3} Z. Yousaf, M. Z. Bhatti, M. F. Malik, Eur. Phys. J. Plus, 
\textbf{134}, 470 (2019).

\bibitem{Yousaf4} Z. Yousaf et al. Eur. Phys. J. C., \textbf{691}, 77 (2017).

\bibitem{6} S. Nojiri, S.D. Odintsov, Phys. Lett. B \textbf{631}, 1 (2005).

\bibitem{7} A. De Felice, S. Tsujikawa, Phys. Rev. D \textbf{80}, 063516
(2009).

\bibitem{8} R. Myrzakulo, Eur. Phys. J. C. \textbf{72}, 2203 (2011).

\bibitem{9} K. Bamba, Eur. Phys. J. C. \textbf{67}, 295 (2010).

\bibitem{11} T. Harko, F.S.N. Lobo, S. Nojiri, S.D. Odintsov, Phys. Rev. D 
\textbf{84}, 024020 (2011).

\bibitem{Yousaf1} M. Z. Bhatti, Z. Yousaf, Zarnoor, Gen. Relat. Gravit., 
\textbf{51}, 144 (2019).

\bibitem{Yousaf2} M. Z. Bhatti, Z. Yousaf, M. Yousaf, Phys. Dark Universe, 
\textbf{28}, 100501 (2020).

\bibitem{Pedro} P.H.R.S. Moraes, P.K. Sahoo, S.K.J. Pacif, Gen. Relat.
Gravit., \textbf{52}, 32 (2020).

\bibitem{Eckart} C. Eckart, Phys. Rev., \textbf{58}, 919 (1940).

\bibitem{Okumura} H. Okumura, F. Yonezawa, Physica A \textbf{321} 207-219
(2003).

\bibitem{Brevik} I. Brevik, O. Gron, J. de Haro, S. D. Odintsov, E.
N. Saridakis, Int. J. Mod. Phys. D \textbf{26}, 1730024 (2017).

\bibitem{Yousaf} M. Sharif, Z. Yousaf, JCAP \textbf{06}, 019 (2014).

\bibitem{20} S. D. Odintsov, Diego Saez-chillon Gomez, G.S. Sharov, Phys.
Rev. D \textbf{101}, 044010 (2020).

\bibitem{10} C. P. Singh, P. Kumar, Eur. Phys. J. C. \textbf{74}, 3070
(2014).

\bibitem{33} W. Misner, Astrophys. J. \textbf{151}, 431 (1968).

\bibitem{34} W. Israel, J.N. Vardalas, Nuovo Cimento Lett. \textbf{4}, 887
(1970).

\bibitem{12} I. Wega, R. C. Falcao, R. Chanda, Phys. Rev. D \textbf{33},
1839 (1986).

\bibitem{27} T. Padmanabhan, S. Chitre, Phys. Lett. A \textbf{120}, 433
(1987).

\bibitem{35} B. Cheng, Phys. Lett. A \textbf{160}, 329 (1991).

\bibitem{36} G. C. Samanta, R. Myrzakulov, Chin. J. Phys. \textbf{55} 1044
(2017).

\bibitem{Davood} S. Davood Sadatian, EPL, \textbf{126}, 30004 (2019).

\bibitem{29} J. Ren, Xin-He Meng, Phys. Lett B \textbf{633}, 1 (2006).

\bibitem{Hiscock} W. A. Hiscock, L. Lindblom, Phys. Rev. D \textbf{31}, 725
(1985).

\bibitem{W. Israel} W. Israel, Ann. Phys. \textbf{100}, 310 (1976).

\bibitem{Stewart} W. Israel, J.M. Stewart, Ann. Phys. \textbf{118}, 341
(1979).

\bibitem{Titus} A. Sasidharan, T. K. Mathew, Eur. Phys. J.C. \textbf{751},
348 (2015).

\bibitem{22} A. G. Riess et al., Astrophys. J. \textbf{607}, 665 (2004).

\bibitem{30} D. J. Eisenstein et al., Astrophys. J. \textbf{633}, 560 (2005).

\bibitem{Fisher} S. B. Fisher, E. D. Carlson, Phys. Rev. D \textbf{100},
064059 (2019).

\bibitem{Tiberiu Harko} T. Harko, P. H.R.S. Moreas, Phys. Rev. D \textbf{101}%
, 108501 (2020).

\bibitem{Setare} M. R. Setare, M. J. S Houndjo, Can. J. Phys. \textbf{91}%
(3), 260-267 (2013).

\bibitem{Rani} M. Sharif, S. Rani, Mod. Phys. Lett. A \textbf{27}, 1350118
(2013).

\bibitem{Iver} I. Brevik, Entropy \textbf{14}, 2302-2310 (2012).

%
%
%

\bibitem{26} A. A. Mamon, S. Das, Eur. Phys. J. C \textbf{77}, 495 (2017).

\bibitem{32} J. R. Garza et al., Eur. Phys. J. C \textbf{79}, 890 (2019).

\bibitem{21} R. A. Knop et al., Astrophys. J. \textbf{598}, 102 (2003).

\bibitem{23} E. E. O. Ishida et al. Astropart. Phys. 28 \textbf{6}, 547
(2008).

\bibitem{24} J. V. Cunha, Phys. Rev. D \textbf{79}, 047301 (2009).

\bibitem{25} N. Rani et al., J. Cosmol. Astropart. Phys. \textbf{1512}, 045
(2015).

\bibitem{visser/2000} M. Visser, C. Barcelo, COSMO-99, 98 (2000).

\bibitem{wald} R. M. Wald, General relativity (University of Chicago Press,
Chicago, 1984).

\bibitem{Hawking/1973} S. W. Hawking, G. F. R. Ellis, The Large Scale
Structure of Space-Time (Cambridge University Press, 1973).

\bibitem{Barcelo/2002} C. Barcelo, M. Visser, Int. J. Mod. Phys. D \textbf{11%
}, 1553 (2002).

\bibitem{moraes/2017} P.H.R.S. Moraes, P.K. Sahoo, Eur. Phys. J. C, \textbf{%
77}, 480 (2017).

\bibitem{Visser/1997} M. Visser, Phys. Rev. D \textbf{56}, 7578 (1997).

\bibitem{43} M.S. Linton et al., J. Cosmol. Astropart. Phys. \textbf{04},
043 (2018).

\bibitem{42} G. Ballesteros, J. Lesgourgues, J. Cosmol. Astropart. Phys. 
\textbf{10}, 014 (2010).

\bibitem{13} V. Sahni, et al. JETP Lett. \textbf{77}, 201 (2003).

\bibitem{15} M. Shahalam, Sasha Sami, Abhineet Agarwal, Mon. Not. R. Astron.
Soc. \textbf{448}, 2948 (2015).

\bibitem{16} V. Sahni, A. Shafieloo, A. A. Starobinsky, Phys. Rev. D \textbf{%
78}, 103502 (2008).

\bibitem{17} C. Zunckel, C. Clarkson, Phys. Rev. Lett. \textbf{101}, 181301
(2008).

\bibitem{41} Jing-Zhao Qi et al., Res. Astron. Astrophys. \textbf{18}, 066
(2018).

\bibitem{18} A. Shafieloo, V. Sahni, A. A. Starobinsky, Phys. Rev. D \textbf{%
80}, 101301 (2009).

\bibitem{31} D. N. Spergel et al., Astrophys. J. Suppl. Ser. \textbf{170},
377 (2007).

\bibitem{sharov} G. S. Sharov, V.O. Vasiliev, Mathematical Modelling and
Geometry \textbf{6}, 1 (2018).

\bibitem{Union2.1 DATA} N. Suzuki et al., Astrophys. J. \textbf{746}, 85
(2012).

\bibitem{ritika/2018} R. Nagpal, S. K. J. Pacif, J. K. Singh1, Kazuharu
Bamba, A. Beesham, Eur. Phys. J. C \textbf{78}, 946 (2018).

\bibitem{gio} R. Giostri et al., J. Cosm. Astropart. Phys., \textbf{1203},
027 (2012).

\bibitem{h14} D. Stern et al., J. Cosmol. Astropart. Phys., \textbf{02}, 008
(2010).

\bibitem{h7} E. Gaztaaga, A. Cabre, L. Hui, Mon. Not. Roy. Astron. Soc., 
\textbf{399}, 1663 (2009).

\bibitem{h13} J. Simon, L. Verde, R. Jimenez, Phys. Rev. D, \textbf{71},
123001 (2005).

\bibitem{h10} A. Oka et al., Mon. Not. Roy. Astron. Soc., \textbf{439}, 2515
(2014).

\bibitem{h6} Y. Wang et al., Mon. Not. Roy. Astron. Soc. \textbf{469}, 3762
(2017).

\bibitem{h16} M. Moresco et al., J. Cosmol. Astropart. Phys., \textbf{08},
006 (2012).

\bibitem{h1} C. H. Chuang, Y. Wang, Mon. Not. Roy. Astron. Soc., \textbf{435}%
, 255 (2013).

\bibitem{h15} C. Zhang et al., Research in Astron. and Astrop., \textbf{14},
1221 (2014).

\bibitem{h11} S. Alam et al., Mon. Not. Roy. Astron. Soc., \textbf{470},
2617 (2017).

\bibitem{h8} C. Blake et al., Mon. Not. Roy. Astron. Soc., \textbf{425}, 405
(2012).

\bibitem{h18} M. Moresco et al., J. Cosmol. Astropart. Phys., \textbf{05},
014 (2016).

\bibitem{h19} A.L. Ratsimbazafy et al., Mon. Not. Roy. Astron. Soc., \textbf{%
467}, 3239 (2017).

\bibitem{h2} C. H. Chuang et al. , Mon. Not. Roy. Astron. Soc., \textbf{433}%
, 3559 (2013).

\bibitem{h5} L. Anderson et al., Mon. Not. Roy. Astron. Soc. , \textbf{441},
24 (2014).

\bibitem{h9} N. G. Busca et al., Astron. Astrop., \textbf{552}, A96 (2013).

\bibitem{h12} J. E. Bautista et al. Astron. Astrophys., \textbf{603}, A12
(2017).

\bibitem{h4} T. Delubac et al., Astron. Astrophys. , \textbf{574}, A59
(2015).

\bibitem{h3} A. Font-Ribera et al., J. Cosmol. Astropart. Phys., \textbf{05}%
, 027 (2014).

\bibitem{h17} M. Moresco, Mon. Not. Roy. Astron. Soc.: Letters. , \textbf{450%
}, L16 (2015).

\bibitem{padn} N. Padmanabhan et al., Mon. Not. Roy. Astron. Soc. \textbf{427%
}, 2132 (2012).

\bibitem{6df} F. Beutler, C. Blake, M. Colless, D. H. Jones, L.
Staveley-Smith, L. Campbell et al., Mon. Not. Roy. Astron. Soc. \textbf{416}%
, 3017 (2011).

\bibitem{boss} BOSS collaboration, L. Anderson et al., Mon. Not. Roy.
Astron. Soc. \textbf{441}, 24 (2014).

\bibitem{wig} C. Blake et al., Mon. Not. Roy. Astron. Soc. \textbf{425}, 405
(2012).

\bibitem{Hz-Plank} P. A. R. Ade et al. [Planck Collaboration], Astron.
Astrophys., \textbf{571}, A16 (2014).

\bibitem{waga} M. Vargas dos Santos, Ribamar R. R. Reis, J. Cosm. Astropart.
Phys., \textbf{1602}, 066 (2016).
\end{thebibliography}
\end{document}